\documentclass[12pt]{article}
\usepackage{graphics}
\usepackage{epsfig}
\usepackage{latexsym}
\newcommand{\beq}{\begin{equation}}
\newcommand{\eeq}{\end{equation}}
\newcommand{\beqa}{\begin{eqnarray}}
\newcommand{\eeqa}{\end{eqnarray}}
\newcommand{\beqar}{\begin{eqnarray*}}
\newcommand{\eeqar}{\end{eqnarray*}}

\newcommand{\eps}{\epsilon}

\newcommand{\ka}{\kappa}

\newcommand{\p}{\phi}

\newcommand{\sss}{\scriptscriptstyle}
\def\sfrac#1#2{{\textstyle{#1\over #2}}}
\newcommand{\eg}{{\it e.g.,}\ }
\newcommand{\ie}{{\it i.e.,}\ }
\newcommand{\labell}[1]{\label{#1}} 
\newcommand{\reef}[1]{(\ref{#1})}

\newcommand\tr{{\tilde r}}
\newcommand\tR{{\tilde R}}
\newcommand\ttheta{{\tilde\theta}}

\def\p1{\phantom{1}}
\def\IR{{\hbox{{\rm I}\kern-.2em\hbox{\rm R}}}}
\begin{document}

\thispagestyle{empty}

\rightline{\small hep-th/0112047 \hfill McGill-01/25}

\rightline{\small \hfill MIT-CTP-3222} \vspace*{2cm}

\begin{center}
{\bf \Huge Dynamical Stability of Six-Dimensional \\ \vspace{0.5cm}
Warped Brane-Worlds}\\[.25em]
\vspace*{1cm}

C.P. Burgess$^{a,}$\footnote{\ cliff@physics.mcgill.ca},
James M. Cline$^{a,}$\footnote{\ jcline@physics.mcgill.ca},
Neil R. Constable$^{b,}$\footnote{\ constabl@lns.mit.edu},
Hassan Firouzjahi$^{a,}$\footnote{\ firouzh@physics.mcgill.ca}
\\

\vspace*{0.2cm}
$^a${\it Department of Physics, McGill University}\\
{\it Montr\'eal, QC, H3A 2T8, Canada}\\

\vspace*{0.2cm}
$^b${\it Center for Theoretical Physics and Laboratory for Nuclear Science} \\
{\it Massachusetts Institute of Technology} \\
{\it Cambridge, MA 02139, USA} \\

\vspace{2cm} ABSTRACT
\end{center}

We study a generalization of the Randall-Sundrum mechanism for generating the
weak/Planck hierarchy, which uses two rather than one warped extra dimension,
and which requires no negative tension branes.  A 4-brane with one 
exponentially large compact dimension plays the role of the Planck brane.  We
investigate the dynamical stability with respect to graviton, graviphoton and
radion modes.  The radion is shown to have a tachyonic instability for certain
models of the 4-brane stress-energy, while it is stable in others, and
massless in a  special case.  If stable, its  mass is in the milli-eV range,
for parameters of the model which solve the hierarchy problem.  The radion is
shown to couple to matter with gravitational strength, so that it is
potentially detectable by submillimeter-range gravity experiments.  The radion
mass can be increased using a bulk scalar field in the manner of Goldberger
and Wise,  but only to order MeV, due to the effect of the large extra
dimension.  The model predicts a natural scale of $10^{13}$ GeV on the
4-brane, making it a natural setting for inflation from the ultraviolet brane.

\vfill \setcounter{page}{0} \setcounter{footnote}{0} \newpage

\section{Introduction} \label{intro}

Brane-world scenarios have undergone considerable theoretical scrutiny
in recent times, largely due to the novel solutions which they provide
for the gauge-hierarchy problem. Three kinds of such proposals have been
made: the ADD scenario  \cite{ADD}, in which two extra dimensions are
very much larger than ordinary microphysical scales; the
intermediate-scale scenario  \cite{BIQK}, in which the scale of gravity
and the compactification scale are very close to one another; and the RS
scenario  \cite{RS}, in which the geometry of the extra dimensions is
exponentially warped.

The RS scenario differs qualitatively from the other two, with its
departure from product-space geometries potentially implying many new
kinds of  low-energy features, including possible violations of
Lorentz-invariance  \cite{LELI}, multiple or quazilocalized gravitons
 \cite{multigrav,kogan} and phenomenologically acceptable  modifications
of the cosmological Friedmann equations  \cite{Fmod},  among other
possibilities.

Most of the effort on exploring  this scenario has focused on 5
spacetime dimensions,  making it difficult to ascertain which
features are generic to warped geometries and which are artifacts of
five dimensions. Indeed, some features of these models are very
likely to be specific to five dimensions. For instance, the
requirement of negative-tension branes -- which are generic in five
dimensions -- does not arise in six  dimensions  \cite{gibbons,dave}.
It is also unlikely that the Friedmann equations are modified in six
dimensions in the same way as in five.  Moreover, the absence of
Kaluza-Klein excitations of the metric's radion mode is also a
5-dimensional artifact, due to the trivial geometrical nature of
1-D manifolds. 

There have been numerous proposals for higher-dimensional generalizations of
the RS idea.   One of the earliest was to consider intersections of
codimension-1 branes as the 3-branes  \cite{intersect}. Others involved
modeling the 3-brane where we are supposed to live, or in some cases where
gravity is localized, as a cosmic string or higher-dimensional defect 
\cite{string}. Other relevant work on warped higher dimensional spaces
includes  \cite{Other6D}. 

A particularly attractive six-dimensional warped model has been considered in
various contexts by several authors  \cite{dave}, \cite{nelson1}--\cite{wilt}.
This model is related to the AdS soliton \cite{positive}, a double analytic
continuation of a planar AdS Black hole metric, and involves two compact
dimensions having the topology of a disc with a conical singularity at its
centre. The boundary of the disc occurs at a (Planck) 4-brane and a (TeV)
3-brane is placed at the conical defect. The stress energy of the 4-brane
requires an anisotropic form which could arise from the smearing of 3-branes
around the 4-brane, as suggested in ref.\ \cite{dave}, or from Casimir energy
of light particles confined to the 4-brane  \cite{nelson1}.

Since all observable consequences of this (or any other) geometry
only involve the theory's low-energy degrees of freedom, essential
to understanding its physical implications is a determination of
its low energy spectrum.  While this has been partially done in
previous references,  it is the purpose of this paper to provide a
complete accounting of the metric modes, especially as regards the
elusive radion mode.

We find results which differ interestingly from what obtains for
five-dimensional RS models.  Instead of a massless radion, we find
that generically the mass squared is nonzero, and possibly
negative, depending on details of the 4-brane stress tensor.  For
a special case involving Casimir energy on the 4-brane, the radion
is an exactly massless modulus at the classical level. Quantum
corrections (which we do not here calculate) might stabilize the
radion in this case. If the radion mode is stable, the magnitude
of the mass squared is of order ($10^{-3}$ eV)$^2$. Both the
stable and loop-stabilized cases might therefore make this mode of
interest for table-top tests of gravity. The tachyonic
instability, if it occurs, does so regardless of the value of the
radial size of the extra dimensions in the static solution. We
show that it is straightforward to cure this problem in the manner
of Goldberger and Wise  \cite{GW}, by adding a bulk scalar field
which couples to the branes.  The mass of the radion, once
stabilized, is suppressed relative to the Planck scale by an
additional fractional power of the warp factor, which puts it in
the MeV rather than TeV range.  Although this could potentially
have been problematic, we find that the coupling of the radion is
similar to that of gravity, because its wave function is not
strongly peaked on the TeV brane.  Therefore, although there are
cosmological contraints on this model, it is not ruled out by
constraints from supernova cooling or radion production in
colliders.

We organize our presentation as follows.  In section 2 we will introduce the
model at the static level.  In section 3 we will find the dynamical
perturbations for 4-D modes which transform as tensors (gravitons),
vectors, and scalars (the radion).  We will show that, whereas the tensor
and vector modes have a massless ground state, the radion mass squared is
generically nonzero and possibly tachyonic,
although its magnitude is exponentially small.  In section 4 we will
show how a bulk scalar field can stabilize the radion mode, and discuss
the phenomenology of the model.  A summary is given in section 5.

\section{The AdS Soliton in Randall-Sundrum Models} \label{review} In this
section we present the AdS soliton \cite{positive}
and its key properties relevant for
braneworld applications. A more detailed description of this spacetime
including  its role in terms of the AdS/CFT correspondence
can be found in ref.\ \cite{positive}.

The $(p+2)$-dimensional AdS Soliton first arose in connection with
the AdS/CFT corespondence~ \cite{juan} as a double-analytic
continuation of a $(p+2)-$ dimensional planar AdS black hole metric.
In this context the AdS soliton geometry is relevant for the strong
coupling description of Lorentzian signature
superconformal field theories in which both
supersymmetry and conformal invariance have been broken~ \cite{edd}.
Specifically for $p=3$ and $p=5$ it provides a description of strongly
coupled $QCD_3$ and $QCD_4$ respectively.

\subsection{The AdS Soliton}

In this article we will be interested in the six dimensional (\ie $p=4$) AdS
soliton for which the line element may be written,
\beq
ds^2 = a(r)\left(f(r)d\theta^2 +  \eta_{\mu\nu}dx^{\mu}dx^{\nu}\right)+
 a^{-1}(r)f^{-1}(r)dr^2    \labell{eq:metric} \eeq
 where the metric functions are given by,
 \beq f(r) = {\rho^2\over L^2}\left(1-\frac{\rho^5}{r^5}\right)\,\,\,\, {\rm
 and} \,\,\,\, a(r)=\frac{r^2}{\rho^2} \label{static} \eeq
 and $\eta_{\mu\nu}$
is the four dimensional Minkowski metric.\footnote{Our metric's signature is
`mostly plus' and we adopt MTW \cite{MTW}
curvature conventions.}\ This is a solution to six
dimensional Einstein gravity with negative cosmological
constant
\beq
\label{Lambdaeq}
\Lambda= -{10\over L^2} \equiv -{8\over 5}k^2,
\eeq
The spacetime is asymptotically locally AdS as
$r\to\infty$; below we will cut off the radial extent by inserting
a 4-brane at a finite value of $r$.  For convenience we have normalized
$a(\rho)=1$, since we will be interested in placing the standard model
on a 3-brane situated at that position.

The range of the $r$ coordinate is $\rho \le r < \infty$, with the
geometry smoothly ending at $r=\rho$ provided that the $\theta$
coordinate is periodic, with period
\beq
\beta=\frac{4\pi L^2}{5\rho}. \labell{period}
\eeq
This also requires that, in the context of supergravity, fermions
are  antiperiodic in the $\theta$ direction. It is important to
note that this  geometry is everywhere smooth and nonsingular
including at $r=\rho$ where the  circle parameterized by $\theta$
smoothly shrinks to a point and the geometry ends. An attractive
feature of this geometry is that it ends in a  natural and
{nonsingular} fashion, allowing constructions similar to those
proposed in refs.\ \cite{singular}, but which are free of
uncontrollable and likely unphysical curvature singularities.

To construct a brane-world model, we will want to  imagine that we
are living on a (TeV) 3-brane at $r=\rho$, thereby introducing a
conical defect there of size $\delta = \kappa^2 \tau_3$, where
$\tau_3$ is the 3-brane tension and $\kappa^2$ is related to the
six-dimensional Newton's constant by $\kappa^2_6 = 8\pi G_6$. This
modifies eq.~(\ref{period}) to become
\beq
\beta=\frac{2 L^2}{5\rho} \; (2\pi - \delta).
\labell{periodd}
\eeq
The extra
dimensions are compactified by terminating the space at a 4-brane at $r=R$.

In the horospheric coordinate system,
the proper distance from $\rho$ to $r$ along the $r$-direction is given by
\beq
\tilde r  \equiv \int_{\rho}^{r} {dr \over \sqrt{a f}} = k^{-1}
\; \cosh^{-1} \left[ \left( {r/\rho} \right)^{5/2} \right],
\eeq
and so $r/\rho = \cosh(2k\tr/5) \sim e^{\tr/L}$ if $r\gg\rho$.

We will often
find it enlightening to express the solution in polar coordinates,
$\tr$, where the line element has the form
\beq
\label{polar}
    ds^2 = a(\tr) \eta_{\mu\nu}dx^{\mu}dx^{\nu} + b(\tr) d\ttheta^2
    + d\tr^2
\eeq
Here the metric coefficients are given by
\beq
    a(\tr) = \cosh^{4/5}(k\tr);\qquad b(\tr) = b_0 {\sinh^2(k\tr)\over
    \cosh^{6/5}(k\tr)}
\eeq
where $b_0 = k^{-2}$ if the point $\tr=0$ is regular, and $\ttheta \in
[0,2\pi]$.  In general we will
suppose the 3-brane has nonvanishing tension located at this point.
Then the conical singularity at $\tr=0$ introduces the deficit
angle given by $2\pi(k\sqrt{b_0}-1)$.  In these coordinates  the proper
distance between two radii $\tr_1$ and $\tr_2$ is simply their difference,
$\tr_2-\tr_1$.  We denote the radial position of the 4-brane by $\tr=\tR$.

We close this section with some comments regarding the stability of the AdS
soliton. Since the AdS soliton is constructed from multiple
analytic continuations of a black hole space time one might worry about
dynamical stability of the solution. In general such
analytically continued space times are not always well behaved. For example
beginning with the Reissner-Nordstrom black hole in asymptotically flat space
one can analytically continue the metric in such a way as to allow for
arbitrarily large negative values for the mass parameter. Solutions such as
this are inherently unstable {\it i.e.} small perturbations  around the
background are tachyonic---see ref.\ \cite{positive} for a  detailed
discussion.

One of the key results of ref.\ \cite{positive} was that (for $p=3$)
the AdS soliton was found to be  perturbatively stable to such
linearized fluctuations.
Further, in ref.\ \cite{spin2} this proof was extended to arbitrary
$p\geq 2$ ---see also ref.\ \cite{wool} for a recent discussion.
One  consequence of this proof is that at least locally within the space of
solutions to the Einstein equations with asymptotically
locally AdS boundary conditions the AdS soliton represents the minimum energy
solution.

It is one of the purposes of this paper to investigate whether
the perturbative stability of this space time persists when the geometry is
cut off by the introduction of the 4-brane discussed above.

\subsection{The Gauge Hierarchy}

To understand how this model solves the gauge hierarchy problem, let us
imagine that all the fundamental scales $M_6$, $k$, and $1/R$ are of order
TeV.  Then the standard reduction of the gravitational action from 6 to
4 dimensions (using polar coordinates) gives the 4-D Planck mass as
\beqa
{M_p^2} &=&  {M_6^4} \int d\tr\,d\ttheta\, a(\tr)\sqrt{b(\tr)}
\nonumber\\ &\sim & {M_6^4\over k^2} a^{3/2}(\tR) =
    {M_6^4\over k^2} e^{6k\tR/5}
\eeqa
Thus by taking $e^{3k\tR/5}\sim 10^{16}$, corresponding to
$k\tR \cong 60$, we can explain the largeness of the Planck scale.

Notice that the relation $M_p^2 \sim a^{3/2}(\tR)$(TeV)$^2$ differs from
the analogous relation in the 5-D RS1 model ({\it i.e.,} the Randall-Sundrum
model which is compactified by the presence of the TeV brane),
$M_p^2 \sim a(\tR)$(TeV)$^2$.  The additional factor of $a^{1/2}$ is
coming from the $b^{1/2}$ part of the measure, which gives the
size of the extra compact dimension that was not present in the 5-D model.
This shows that the present model is a hybrid of the RS and ADD scenarios,
in that the hierarchy is due to a combination of warping and having a
large extra dimension.

This difference can also be seen by considering the physical mass of a
4-D scalar field which is confined to a 3-brane at a fixed position
$(\tr,\ttheta)$ in the bulk.  Since we are taking the fundamental scale to
be TeV, we should assume that its bare mass parameter $m$ is of this order.
But the physical mass is determined by the usual argument of canonically
normalizing its kinetic term:
\beqa S_3 &=& - \, \frac12 \int d^4x \; a^2 \, \Bigl[
a^{-1}\eta^{\mu\nu} \partial_\mu \phi \partial_\nu \phi +
m^2\phi^2 \Bigr]\nonumber\\ &\to& - \, \frac12 \int d^4x \;
\Bigl[\eta^{\mu\nu} \partial_\mu \phi \partial_\nu \phi +
a(\tr)m^2\phi^2 \Bigr] \eeqa
Thus the physical mass is given by $m_3 = m \sqrt{a(\tr)}$.  If we
take the particle all the way to the 4-brane, its mass does not
reach the Planck scale, but rather a smaller one,
$a(\tR)^{-1/4}M_p \sim 10^{-32/6} M_p\sim 10^{13}$ GeV.  This
reflects the fact that the strength of gravity is still diluted
for a 4-D observer on the 4-brane, by the large extra dimension.
We should refer to it as the ``$10^{13}$ GeV brane'' rather than
the Planck brane.

\subsection{Properties of the Branes}

The 4-brane we need for compactifying the
AdS soliton can be constructed by the standard cutting and
pasting procedure. Here, the metric in eq.\ \reef{static} will be cut along the
surface $r=R$ and then pasted onto a mirror image of itself. The resulting
space time is then a solution of
\beq
\kappa^{-2}{\sqrt{-G}}\left(R_{MN}-\frac{1}{2}RG_{MN}-
\Lambda G_{MN}\right) = \sqrt{-g}S_{ab}\delta^{a}_{M}\delta^{b}_{N}
\delta\left({r-R\over\sqrt{fa}}\right),
\labell{bigeqn}
\eeq
where $G_{MN}$ is the metric given in eqn.\reef{static}, $\Lambda=-10L^{-2}$
is the cosmological constant appropriate to six dimensional AdS spaces,
$\ka^{2}=8\pi G_{6}$, and the induced  metric on the 4-brane is
$g_{ab}= G_{MN}(R)\delta^{M}_{a} \delta^{N}_{b}$.

$S_{ab}$ is the stress tensor of an infinitely thin
brane located at the cutting surface. The stress-tensor so defined may be
obtained  from the Israel matching conditions~ \cite{Israel}. A straightforward
 calculation yields,

\beqa
\labell{jump_cond1}
S_{\mu\nu}&=&\kappa^{-2}\left(4\frac{a^{\prime}}{a}+
\frac{f^{\prime}}{f}\right)g_{\mu\nu}
\\
S_{\theta\theta}&=&
\kappa^{-2}\left(4\frac{a^{\prime}}{a}\right)g_{\theta\theta}.
\labell{jump_cond2}
\eeqa
Here and in the following uppercase latin indices indicate six dimensional
coordinates,  while lower case greek indices specify the
coordinates parameterizing the directions along the 3-brane. Lower-case
latin indices similarly label directions parallel to the 4-brane
\ie $x^a=(x^{\mu},\theta)$.

A crucial point for this model is that the extra term $f'/f$ in eq.\
(\ref{jump_cond1}) relative to (\ref{jump_cond2}), though small, is
nonvanishing, and therefore it is impossible to interpret the 4-brane stress
tensor as being due to a pure tension.  Were we to do this, thus making
$S_0^{\ 0} = S_{\theta}^{\ \theta}$, the 4-brane would be forced to go to
$r=\infty$, and we would lose the compactification of the extra dimensions
and the localization of gravity.

There are several kinds of physics on the 4-brane which would
naturally involve the required difference in $S_0^{\ 0} -
S_{\theta}^{\ \theta}$. One is to imagine that the gluing surface
is composed of multiple branes. As discussed in ref.\ \cite{dave},
one could consider the superposition of the stress-energy tensors
of a four-brane wrapping the internal circle and a three-brane
which is smeared over the internal circle. Indeed eqs.\
(\ref{jump_cond1},\ref{jump_cond2}) then take the
form~ \cite{dave},
\beqa
S_{\mu\nu}&=&\left(T_4+\frac{T_3}{L_{\theta}}\right)g_{\mu\nu}
\nonumber \\
S_{\theta\theta}&=&T_4\,g_{\theta\theta}
\labell{rob}
\eeqa
where $L_{\theta}$ is the proper period of the circle
parameterized by $\theta$ at $r = R$.

Another very physical possibility is that the difference between
$T_0$ and $T_\theta$ is due to the Casimir energy of any massless
fields which are confined to the 4-brane  \cite{nelson1}. For these
the stress-energy tensor will take the form
\beqa
S_{\mu\nu}&=&\left(T_4+\frac{c_0}{L_{\theta}^5}\right)g_{\mu\nu}
\nonumber \\
S_{\theta\theta}&=&\left(T_4-\frac{c_\theta}{L_{\theta}^5}\right)
g_{\theta\theta}
\labell{casimir}
\eeqa
with some dimensionless coefficients $c_0$ and $c_\theta$. To the
extent that the trace anomaly vanishes (which is the case at one loop, since
the 4-brane is odd-dimensional), the Casimir energy satisfies the
condition $g^{ab}S_{ab} = 0$, which implies $c_\theta = 4 c_0$.

In a static background, either of these stress-energy tensors are trivially
conserved on the 4-brane.  But when we discuss dynamical perturbations  of
the static space, conservation of stress-energy will yield a nontrivial
constraint on the components of $S_{\mu\nu}$. This will be discussed in
section 3.3, where we  show that the ground state of the radion modes is
tachyonic for the smeared 3-brane model, but massless for the Casimir model.

The issue of stabilization is closely related to another potential
problem with the above models.  This concerns the order of
magnitude of the difference $S_0^{\ 0} - S_{\theta}^{\ \theta}$,
which is required in order to acheive $a(R)\sim 10^{21}$ as is
needed to solve the hierarchy problem. This requires
  \beq
    {S_0^{\ 0} - S_{\theta}^{\ \theta}\over
    S_0^{\ 0} + S_{\theta}^{\ \theta}}
 \sim a(R)^{-5/2} \sim 10^{-53},
\eeq
which appears to be extremely fine-tuned.
From this standpoint, only the Casimir effect can be considered to be
natural, since its $L_\theta^{-5}$ dependence scales precisely like
$a(R)^{-5/2}$.  However the Goldberger-Wise stabilizing field makes it
unnecessary to have nonvanishing $S_0^{\ 0} - S_{\theta}^{\ \theta}$,
 as was shown by  \cite{nelson2}: with the scalar it becomes possible
to achieve an exponential hierarchy even when $S_0^{\ 0} =
S_{\theta}^{\ \theta}$.  It is interesting that we are able to
both determine the size of the extra dimension and stabilize the
radion using the same scalar field.  In the 5D RS1 model, the two
phenomena are necessarily tied together, but not so in 6D.  The
fact that the size of the extra dimension is determined by the
value of $S_0^{\ 0} - S_{\theta}^{\ \theta}$ does not prevent the
instability we will demonstrate, so it is not obvious that
introducing a new effect to determine the size of $R$ should
stabilize the system.  Nevertheless we shall show that it is true.

\section{Stability Analysis}

In the original model of Randall and Sundrum with two branes, fluctuations of
the metric were  decomposable into Kaluza-Klein modes. Most notably the
spectrum included a zero mode which was bound to the brane and served as the
graviton in a four dimensional world. The remaining excitations formed  a
tower of massive modes which  were fully five dimensional and had very little
support near the brane. In the AdS soliton model presented here we will find a
very similar story emerging with a few differences. As in RSI, the spacetime
constructed in the previous section is finite and one can view the graviton
fluctuations as  linearized gravity in a box. This implies that the spectrum
of gravitons will again be discrete.
Another difference from RSI is that the fluctuations of the
metric are now more complicated owing to the greater
complexity of the background metric.
With only a single extra dimension, the only degree of freedom
for the radion mode is the distance between the two branes, since any apparent
ripples in the $dr^2$ metric component can be gauged away by a coordinate
transformation.  With two or more extra dimensions this is no longer the case,
and the radion too has a KK tower of excitations.

By virtue of the symmetries of the geometry at least four of the
metric modes must be exacly massless. First, there are two
massless states which correspond to the massless 4-D graviton
which is ensured by the model's unbroken Lorentz invariance in the
directions parallel to the 3-brane. Second there are two states
making up the massless 4-D spin-one particle, which is a
consequence of the isometry $\theta \to \theta + c$ of the extra
dimensions.

The counting of massless modes may be further sharpened as
follows. If gravity is indeed localized on the 4-brane, we would
expect to find a total of five zero modes appropriate to the five
independent fluctuations of a massless spin-two particle in $4+1$
dimensions. Since we will find below that the radion generically
has a nonvanishing mass in this theory (either tachyonic or real
by adding the appropriate scalar), there are in fact only four
zero modes bound to the brane.  These will have a natural
interpretation, at energies below the mass of the radion, as a
$3+1$ dimensional graviton and a $3+1$ dimensional massless vector
field.

The analysis will proceed by linearizing eq.\ \reef{bigeqn} around the
background of the AdS soliton. In particular we will consider fluctuations of
the six dimensional metric which are given by $g_{MN}\rightarrow g_{MN}+
h_{MN}$.  A feature here is the fact that $h_{MN}$ is a tensor and there is
thus a variety of polarizations, or graviton modes,  that need to be
considered.  Following refs.\ \cite{spin2,brower} we can divide the various
polarizations of the six dimensional graviton into three categories. {\bf (i)}
Transverse traceless polarizations. These are modes which are polarized in
directions parallel to the Lorentz invariant hypersurface spanned by the
coordinates $x^{i}$. {\bf (ii)}  Vector polarizations. These are gravitons
whose polarization is of the form  $\epsilon_{i\theta}$, {\it i.e.,} modes
polarized along the circle and in the flat piece of the brane. {\bf (iii)}
Scalar Polarizations. These are modes which are diagonal but not traceless. It
is this last mode which is related to the radion field.

For the cases {\bf (i)} and {\bf (ii)} above we may write the
metric fluctuations as $h_{MN}= H_{MN}(r)e^{ik\cdot x}$ where
$H_{MN}(r)$ is the radial profile tensor and $k^\mu$ is a
$4$-dimensional momentum vector with $k^2= \eta^{\mu\nu}k_\mu
k_\nu = -M^2$. Further there are ambiguities in the metric
perturbations arising from diffeomorphism invariance, which we
(partially) handle by imposing a ``transverse gauge:''
$H_{M\mu}k^\mu = 0$. For massive excitations we may always, via
the appropriate Lorentz boost, choose to work in the rest
frame\footnote{Of course when searching for zero modes we are
constrained to work with a null momentum vector.} so that the
momentum can be written as $k^\mu=\rho\,\delta^\mu_{t}$. In this
case, the transversality condition becomes,
\beq H_{a\mu}k^{\mu} = 0 \;\;\;\Rightarrow \;\;\;H_{a t}=0
\;\;\;\;\forall a \labell{trans} \eeq
Our implicit notation for the $3+1$-dimensional Minkowski
space coordinates is $x^\mu=(t,x^i)$ with $i=1,\ldots,3$.

We do not consider here the Kaluza-Klein modes\footnote{For a
discussion of these KK modes as they relate to the stability of
the AdS soliton see ref.\ \cite{spin2}.} corresponding to angular
excitations, {\it i.e.}, around the large extra dimension. One
might at first have thought that these had a mass gap of order 10 eV
since they are modes which are localized on the 4-brane (assuming
they are radially unexcited) and the circumference of the 4-brane is of
order $L_\theta\sim (10$ eV$)^{-1}$.  However, if we imagine 
integrating out only the radial dimension to obtain the effective
theory of these modes, we find that the fundamental scale is no
longer TeV, but rather $\sqrt{a(R)}\,$TeV $\sim 10^{13}$ GeV,
because of the effect of warping.  (The kinetic term of these
excitations gets the same exponential factors as does the angular
gradient term: ${\cal L}\sim (\dot\phi)^2/a + (\partial_\theta\phi)^2/b$.) This
effect of warping was alluded to in section 2.2.  From this point
of view, the size of the compact dimension looks like
(TeV)$^{-1}$, whose smallness compared to $\sim 10^{13}$ GeV is
how the largeness of the extra dimension is manifested.  We will
leave aside these angular KK modes and instead consider the radial
excitations. To determine the spectrum, we must solve
eq.~\reef{bigeqn} with the ansatz (\ref{trans}) as an eigenvalue
problem for the mass $M$.

The metric fluctuations for case ({\bf iii}) are considerably more
complicated and will be dealt with separately in section 3.3.

\subsection{Transverse Traceless Modes}

As explained above, these modes are polarized parallel to the Lorentz invariant
directions on the brane and correpond to
\beq
H_{\theta M}  =  H_{r M}  =  0 = H_{M \mu}k^{\mu} \qquad\forall\, M\ ,
\labell{pol1}
\eeq
where the last equality is a restatement of eq.\ \reef{trans}.

A consistent solution to eq.\ \reef{bigeqn} linearized around the AdS soliton
is provided by the following ansatz,
\beq
H_{MN}=\varepsilon_{\sss MN}a(r)H(r)
\labell{ttansatz}
\eeq
where $\varepsilon_{\sss MN}$ satisfies the conditions in eq.\ \reef{pol1} and
$a(r)$ is the metric function appearing in the static solution above.

Solving the equations of motion, which come from linearizing eqn.
\reef{bigeqn} around our ansatz, imposes that the polarization,
must also be traceless,
\beq \eta^{\mu\nu}\varepsilon_{\mu\nu}=0\
. \labell{traceless} \eeq
Thus eq.~\reef{pol1} describes five
independent modes, which can be described as three off-diagonal
polarizations,
\eg \beq \varepsilon_{12}=\varepsilon_{21}=1\
,\qquad {\rm otherwise}\ \varepsilon_{\sss MN}=0 \labell{exampoff}
\eeq
and two traceless diagonal polarizations,
\eg \beq
\varepsilon_{11}=-\varepsilon_{22}=1\ ,\qquad {\rm otherwise}\
\varepsilon_{\sss MN}=0\ . \labell{exampdia} \eeq
For all of these independent polarizations, the radial profile
$H(r)$ satisfies the same differential equation. Substituting the
above ans\"atze, \reef{pol1} and \reef{ttansatz}, into
eq.~\reef{bigeqn} and linearizing around the AdS soliton
background one finds that the radial profile must satisfy
\beq \frac{d^2H(r)}{dr^2}+\left(3\frac{a^{\prime}(r)}{a(r)}+
\frac{f^{\prime}(r)}{f(r)}\right)\frac{dH(r)}{dr}+\frac{M^2}{f(r)a(r)^2}H(r)=0
\labell{tteqn} \eeq
where primes denote differentiation with
respect to $r$.  Here the $\delta$-function coming from the right
hand side has been canceled  exactly by similar terms on the left
hand side. One should note that this is {\it exactly} the equation
for the transverse traceless modes originally obtained in
ref.\ \cite{spin2,brower}---see also ref.\ \cite{kogan}.  Again,
this is independent of the form we choose for the stress tensor,
since by exciting these transverse traceless modes we are not
perturbing the size of the circle. It is interesting to note that
this is precisely the equation describing the propagation of a
minimally coupled massless scalar on the AdS soliton
background~ \cite{spin2,brower}.

Here we will determine the eigenvalues numerically using a shooting technique
(see ref.\ \cite{recipe}). For the purposes of numerical calculation we will
henceforth restrict to $r<R$ and replace the brane by effective boundary
conditions on the gravitons at $r=R$.  Obtaining the correct solution to
this problem is  equivalent  to determining the correct boundary conditions
that the radial profile $H(r)$ must obey at $r=\rho$ and $r=R$. At the brane
the absence of $\delta$-functions in eq.\ \reef{tteqn} implies that $H(r)$ and
its first derivative are continuous so requiring an even function of $r$
gives the boundary condition $H^{\prime}(r)=0$. The  boundary at $r=\rho$ is
more subtle since the metric function $f(r)$  vanishes there, {\it i.e.,}
$f(r=\rho)=0$.  This is reflected in the fact that this point is a regular
singular point of  eq.\ \reef{tteqn}. So requiring that $H(r)$ be regular at
this boundary gives the condition
\beq
\left.\frac{dH(r)}{dr}\right|_{r=\rho}= -\frac{L^2M^2}{5}\left.
H(r)\right|_{r=\rho}
\labell{ttbound}
\eeq
It is now straightforward to see that this polarization of the six dimensional graviton
indeed has a zero mode. This can be done by setting $M^2=0$ in eq.\ \reef{tteqn} and
integrating directly. After the first integration we find
\beq
\frac{dH(r)}{dr}={\rm const.}\times a(r)^{-3}f(r)^{-1}
\labell{bad}
\eeq
which can be seen by inspection to violate the boundary condition at the brane
 unless the integration constant is forced to vanish. Performing the second
integration just leaves the constant solution obeying the above
boundary conditions. So we see, referring back to the ansatz in
eq.\ \reef{ttansatz}, that the physical zero mode for these polarizations is
\beq
h_{MN}=\varepsilon_{\sss MN}a(r)
\labell{zerowave}
\eeq
where $\varepsilon_{\sss MN}$ obeys the conditions in eq.\ \reef{ttansatz} and
we have used our freedom to perform one overall rescaling of the solution
to set $H(r)=1$.

In order to obtain the spectrum of nonzero modes we use the
numerical shooting technique with the above boundary conditions.
The mass eigenvalues are a function of the relative size of the
extra dimension $R/\rho$, but in the limit that $R/\rho$ becomes
exponentially large, as desired to solve the hierarchy problem,
the masses quickly approach their asymptotic values.  We give
these values for the first few KK modes in the following table. We
emphasize that there are no modes with $M^2<0$ and hence no
instabilities in this sector.

\begin{center}
\begin{tabular}{|c|c|} \hline
Mode Number & $ M^2 L^2 $ \\   \hline
0 & 0    \\ \hline
1 & \p1 16.494  \\ \hline
2 & \p1 44.731  \\ \hline
3 & \p1 85.545  \\ \hline
4 & 138.92  \\ \hline
5 & 204.85 \\ \hline
\end{tabular}
\end{center}
\noindent{\small
Table 1. Mass squared of the radial graviton KK modes, in units of the
AdS curvature radius, in the limit of large warp factor.}\\

\indent Unlike the zero mode, which is localized on the 4-brane, the
KK modes are peaked at the TeV brane, a phenomenon which is familiar from
the 5D RS model.  This behavior is illustrated for the first three modes
in figure 1, where the wave functions are plotted.
\centerline{\epsfysize=2.5in\epsfbox{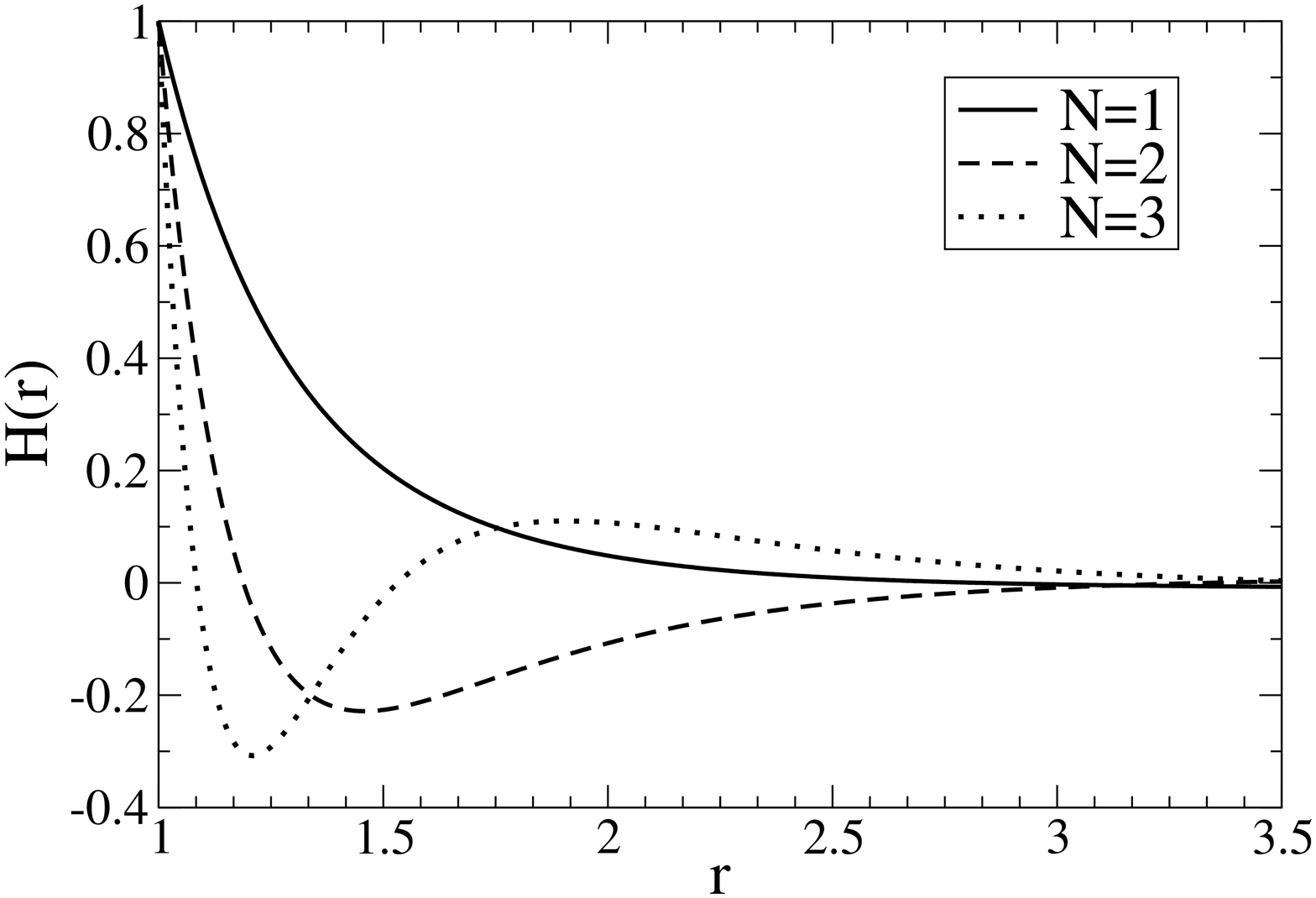}}
\centerline{\small
Figure 1. Wave functions for the first three radial KK modes of the
graviton.}

\subsection{Vector Modes}

The next set of polarizations comes from the same ansatz as in eq.\
\reef{ttansatz}; however in this case the polarization tensor is such that it
has one leg in the lorentz invariant directions of the four-brane and another
on the circle. It has the form
\beq
\varepsilon_{\theta\mu}=\varepsilon_{\mu\theta}=v_{\mu}
\qquad{\rm with}\ k\cdot v=0\ {\rm and}\ v\cdot v=1
\labell{vecansatz}
\eeq
Polarizations of this form contain three independent modes. Substitution
into the equation of motion~\reef{bigeqn} and linearizing as before yields
\beq
\frac{d^2H(r)}{dr^2}+3\frac{a^{\prime}(r)}{a(r)}\frac{dH(r)}{dr}
+\frac{M^2}{f(r)a(r)^2}H(r)=-2\frac{f^{\prime}(r)}{f(r)}\delta(r-R)H(r)
\labell{veceqn}
\eeq
For these modes there is a net contribution from the $\delta$-function source
term on the right hand side of eq.\ \reef{bigeqn}.  This can be understood
from the fact that the metric perturbation we are considering is a fluctuation
in the $g_{\theta x}$ components of the metric. The  corresponding variation
of the stress-energy tensor only involves the pieces
proportional to $T_4$ and not those proportional to $T_3$. In other words for
these modes we have {\it effectively} that $\delta T_4 = \delta T_0$ under the
fluctuations condsidered in this section. In our formalism this will manifest
itself as a nontrivial boundary condition at $r=R$,
\beq
\left.\frac{dH(r)}{dr}\right|_{r=R}=\left.\frac{f^{\prime}(r)}
{f(r)}H(r)\right|_{r=R}
\labell{vecbound}
\eeq
while enforcing regularity at the singular point $r=\rho$ requires that
$H(\rho)=0$. As with the transverse traceless modes of the previous section,
we can obtain an analytic solution for the zero mode of this equation by
setting $M^2=0$ and performing the integration directly. We find an exact
solution of the form
\beq
H(r)=-\frac{c}{5}\frac{L^6}{r^5}+b
\eeq
where $c$ and $b$ are arbitrary (dimensionful) constants of integration.
Imposing the boundary condition in eq.\ \reef{vecbound} gives
%
%
\beq b=\frac{c\, L^6}{5\rho^5} 
\labell{bee} \eeq
and from this it is straightforward to see that the zero mode is
given by
\beq H(r)= \frac{c L^8}{5 \rho^7} \, f(r) 
\labell{veczero} \eeq
Fortuitously this solution also satisfies the boundary condition
at $r=\rho$ for all values of $c$ since this is precisely where
$f(r)=0$. Choosing to normalize the wavefunction so that $H(R)=1$
amounts to choosing $c=5\frac{\rho^7}{L^8}f(R)^{-1}$. Returning to
the ansatz in eq.\ \reef{vecansatz} we see that the physical zero
mode takes the simple form \beq
H_{\mu\tau}=\varepsilon_{\mu\tau}a(r)\frac{f(r)}{f(R)}
\labell{vecphys} \eeq which is indeed peaked at $r=R$. In order to
analyze the nonzero modes we again turn to numerics and find a
positive definite spectrum which implies that no instabilities are
caused by exciting these vector modes. In the table below we again
present, in the limit of large warp factor, the first few
eigenvalues in this spectrum.

\indent Like the spin-2 modes, the vector
KK modes are also peaked near the TeV brane, although their wave function
vanishes precisely there.
The first three modes are shown in
figure 2.

\begin{center}
\begin{tabular}{|c|c|} \hline
Mode Number & $ M^2 L^2$ \\   \hline
0 & 0  \\ \hline
1 & \p1 25.001 \\ \hline
2 & \p1 59.752 \\ \hline
3 & 106.91  \\ \hline
4 & 166.61   \\ \hline
5 & 238.80  \\ \hline
\end{tabular}
\end{center}
\noindent{\small
Table 2. Mass squared of the radial graviphoton KK modes, in units of the
AdS curvature radius, in the limit of large warp factor.}\\

\centerline{\epsfysize=2.5in\epsfbox{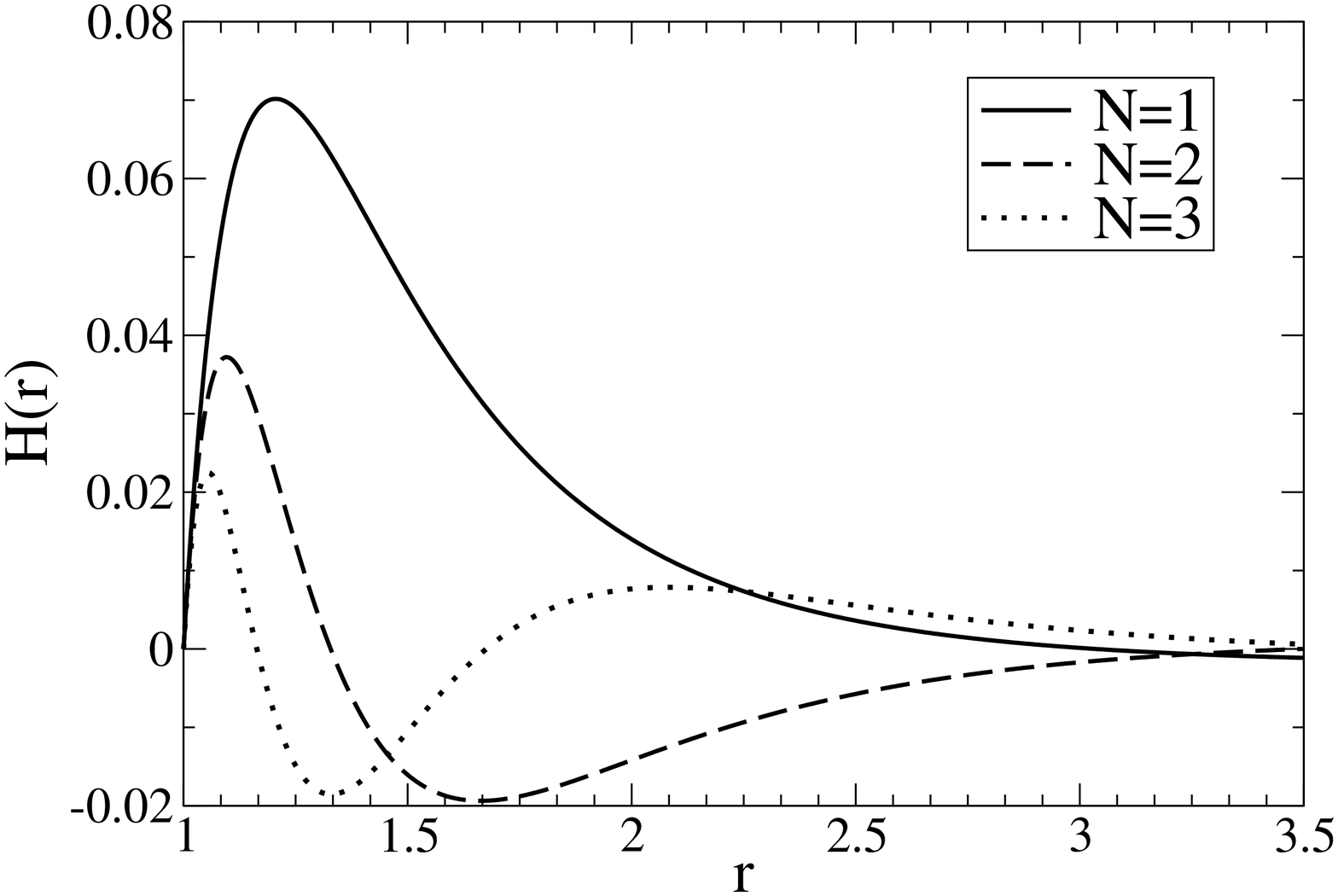}}
\noindent{\small
Figure 2. Wave functions for the first three radial KK modes of the
graviphoton.}

\subsection{Scalar Modes}

Here we will consider modes which fall into category {\bf (iii)} above. There
is in fact only a single mode (at the lowest KK level),  although it
corresponds to simultaneous fluctuations of different components of the
metric. Most importantly this mode includes fluctuations of both
$g_{\theta\theta}$ and  $g_{rr}$ and thus couples fluctuations of the radial
size to fluctuations in the size  of the compact circle.  For the radion, it
is convenient to switch to the polar coordinates introduced in eq.\
(\ref{polar}), where the equations take a somewhat simpler form. We start by
writing the perturbed metric as
\beq
 ds^{2}=a(\tr,t)\eta_{\mu \nu}dx^{\mu}dx^{\nu}+b(\tr,t)d\ttheta^{2}+
c(\tr,t)d\tr^{2}
\labell{pertmet}
\eeq
where the perturbation around the static background
corresponds to
\beqa
a(\tr,t)&=& e^{-A_0(r) - A_1(r,t)}  \equiv a_0(\tr)e^{ -A_1(\tr,t)}
\nonumber \\
b(\tr,t)&=&  e^{-B_0(\tr) - B_1(\tr,t)} \equiv  b_0(\tr) e^{ -B_1(\tr,t)}
\nonumber \\
c(\tr,t)&=&  1+C_1(\tr,t)
\labell{stat}
\eeqa
and we have now included a subscript `$0$' to denote the  metric functions of
the background around which we are expanding, \ie the AdS soliton.  Here we
have gone to the 4-D rest frame of the fluctuation
\beq
    A_1(\tr,t) = A_1(\tr) {\rm Re} (e^{-i\omega_r t})
\eeq
and similarly for $F_1$ and $G_1$.  This is generically valid since for arbitrary
values of the parameters there will be no zero mode. We will find however that
there is a special case in which there is a zero mode and hence no rest frame.
For this case we have checked that the procedure outlined here gives
the correct result.

We can unify the models of the 4-brane stress energy at $\tr=\tR$
which were discussed in section 2.3 by writing
\beqa
S_{\mu\nu}&=&\left(T_4+\frac{T_3}{{L_{\theta}}^\alpha}\right)g_{\mu\nu}
\equiv  V_0\, g_{\mu\nu}\nonumber \\
S_{\theta\theta}&=&\left(T_4-\frac{T_3'}{{L_{\theta}}^\alpha}\right)
g_{\theta\theta} \equiv V_\theta\, g_{\theta\theta}
\labell{rob2}
\eeqa
where $L_\theta = \int d\theta \sqrt{b}|_{\tr=\tR}$ is the circumference of
the compact 4-brane dimension.   The smeared 3-brane model has $\alpha =1$
and $T_3'=0$, while the Casimir effect model has $\alpha = 5$ and
$T_3,T_3'\neq 0$.  Notice that
the metric tensor elements appear within the definition of the circumference
$L_\theta$.  Also $T_3$ really has the dimensions of a 3-brane tension only when
$\alpha=1$.

We can now write the Einstein equations for this metric ansatz.
Since it will later be necessary to add a bulk minimally-coupled
scalar field, we include it here, although at first we shall carry
out the analysis with no scalar. Ignoring terms like $\dot a^2$
which would be higher order in the perturbations, the ($tt$),
($tt)+(ii$), ($rr$), ($\theta\theta$) and ($tr$) components of the
Einstein equations are
\beqa \label{tt}
      \frac{3}{2}\frac{a^{\prime \prime}}{a}
      +\frac{3}{4}\frac{a'}{a}\frac{b'}{b}
     -\frac{3}{4}\frac{a'}{a}\frac{c'}{c}\!\!\!&\!+&\!\!\!\!
     \frac{1}{2}\frac{b^{\prime \prime}}{b}-
    \frac{1}{4}\left(\frac{b'}{b}
    \right)^{2}
     -\frac{1}{4}\frac{b'}{b}\frac{c'}{c} +
    \nonumber\\
  &=&-\kappa^{2}\left(c[\Lambda+V(\phi)] + \sfrac12\phi'^2
     +V_{0}\sqrt{c}\delta(\tr-\tR) 
    + \ {T_b}\sqrt{\frac{c}{b}}\delta(\tr)\right)
\eeqa
\vspace{-0.55cm}
\beqa
\label{ttxx}
     2\frac{\ddot{a}}{a}+\frac{\ddot{b}}{b}
    +\frac{\ddot{c}}{c}&=&0\\
\label{rr}
    \frac{3}{2}\left(\frac{a'}{a}\right)^{2}+\frac{a'}{a}\frac{b'}{b}
   -\frac{1}{2a_{0}}\left(\frac{\ddot{b}}{b}
    +3\frac{\ddot{a}}{a}\right)
    &=&-\kappa^{2}\left(c[\Lambda+V(\phi)] -\sfrac12\phi'^2\right)\\
\label{thth}
    2\frac{a^{\prime \prime}}{a}+\frac{1}{2}\left(\frac{a'}{a}\right)^{2}
    -\frac{a'}{a}\frac{c'}{c}-\frac{1}{2a_{0}}
\left(3\frac{\ddot{a}}{a}
   +\frac{\ddot{c}}{c}\right)&=&-\kappa^{2}\left(c[\Lambda+V(\phi)]
   + \sfrac12\phi'^2+V_{\theta}\sqrt{c}\delta(\tr-\tR)\right)
\nonumber \\ \\
\label{tr}
   6\frac{\dot{a'}}{a}-6\frac{\dot{a}}{a}\frac{a'}{a}
   -\frac{\dot{b}}{b}\frac{a'}{a}-3\frac{\dot{c}}{c}\frac{a'}{a}
   +2\frac{\dot{b'}}{b}
   -\frac{\dot{b}}{b}\frac{b'}{b}-\frac{\dot{c}}{c}\frac{b'}{b}&=&
    4\kappa^2 \dot\phi \phi'
\eeqa
where $T_b$ is the tension of the 3-brane at $\tr=0$
and primes denote $\partial_\tr$.

The next step is to linearize the field equations in the dynamical
perturbations.
We use the relation $\ddot A_1 = - m^2_r A_1$, and similarly for the
other perturbations, where $m_r$ is the
sought-for radion mass.
Expanding the $(tt)+(ii)$, $(tr)$ and $(rr)$
Einstein equations, respectively, to first order gives
\beq
\label{(tt-ii)}
       m^2_r (2A_{1}+B_{1}-C_{1})=0
\eeq
\beq
\label{(tr)}
    4(A_{1}'-\frac{1}{2}A_{0}'C_{1})
    +\left[B_{1}'-A_{1}'-\frac{1}{2}(B_{0}'-A_{0}')(B_{1}+C _{1})\right]
    -2\kappa^{2}\phi_{0}'\phi_{1}=0
\eeq
\beqa
\label{(rr)}
    (3A_{0}'+B_{0}')A_{1}'+A_{0}'B_{1}'-A_{0}'
    \left(\frac32 A_{0}'+B_{0}'\right)C_{1}
    -\frac{m_{r}^{2}}{2a_{0}}(3A_{1}+B_{1})\nonumber\\
    +{\kappa^{2} \over 2}\left[{\phi_{0}'}^{2}C_{1}+
    2\frac{\partial V}{\partial \phi}\phi_{1}-2\phi_{0}'\phi_{1}'
    \right]=0
\eeqa
and similarly for the scalar field we find
\beqa
\label{scalar}
       \phi_{1}''-(2A_{0}'+B_{0}')\phi_{1}'
       -\frac{1}{2}(4A_{1}'+B_{1}'
       +C_{1}')\phi_{0}'-\left(C_{1}\frac{\partial V}{\partial \phi}
       +\frac{\partial^{2}V}
       {\partial \phi^{2}}\phi_{1}
       -\frac{m_{r}^{2}}{a_{0}}\phi_{1}\right)=0.
\eeqa
We note that even for a massless mode, the energy is nonvanishing, so that
eq.\ (\ref{(tt-ii)}) would still provide a constraint among the components of the
perturbation (we would then have to consider its spatial momentum too).
Moreover (\ref{(tr)}) is the time-integrated form of the $(tr)$ equation,
which we have organized in a form whose purpose will become clear momentarily.
We have not written the $(tt)$ nor $(\theta\theta)$ components since these
are not independent equations: $(tt)$ can be obtained from $(tr)'$ and
$(tt)+(ii)$, which follows from the Bianchi identities; and $(\ttheta\ttheta)$ follows from combining $(tr)$, $(rr)$,
$(rr)'$ and $\phi$ equations in a manner which is not obvious, but which
could be anticipated, since first order constraint equations, like $(rr)$
and $(tr)$, must be consistent with the second-order dynamical equations,
$(tt)$, $(ii)$ and $(\ttheta\ttheta)$.

For the remainder of this section, we will assume there is no bulk scalar
field present.  Effects of the bulk scalar will be considered in the next
section.

The boundary conditions at the 4-brane come from integrating over the delta
functions to find the discontinuity in the first derivatives, and using
$Z_2$ symmetry across the brane to identify, {\it e.g.,} $\Delta a'/a =
-2a'/a$ at $\tr=\tR$.  At zeroth order in the perturbations, this gives the
jump conditions (\ref{jump_cond1},\ref{jump_cond2}) for the static solution.  Expanding to
first order, and assuming that $T_3$ is a constant
(more about this below) we find
\beqa
\label{bcA}
    A_{1}'&=& \frac{1}{2}A_{0}'C_{1} - {\alpha\over 8}\left(T_3'\over
T_3+T_3'\right)
    (B_{0}'-A_{0}')B_{1}
     \\ 
\label{bcB}
    B_{1}'-A_{1}' &=& \frac{1}{2}(B_{0}'-A_{0}')\left(C_{1}+
    \alpha B_{1}\right).
\eeqa
In addition, we must consider the boundary condition at the 3-brane at the
center, $\tr=0$.  With two extra dimensions, the effect of such a point-like
source is to introduce a conical singularity and a corresponding deficit angle,
as we have discussed in section 2.1.  Since the defect is unchanged by
perturbations around the static solution, we insist that the deficit angle
does not vary.  If we consider a circle with $\tr=\epsilon$ around the 3-brane,
with circumference $L$ and physical radius $D$, we therefore demand that
$L/D$ be invariant in the limit that $\eps\to 0$:
\beqa
    \lim_{\eps\to 0}\, \delta \left({L\over D}\right) &=&
    \lim_{\eps\to 0}\, \delta \left({\int d\ttheta\sqrt{b}\over \int_0^\epsilon
    \sqrt{c}}\right)
    \nonumber\\
    &=& -{L\over 2D}\left.(B_1+C_1)\right|_{\tr=0},
\eeqa
in other words, $B_1+C_1=0$ at $\tr = 0$.  In addition, we expect
$A_1'$ to vanish at $\tr=0$.  The latter can be shown to be
satisfied from $B_1+C_1=0$ combined with the bulk equations of
motion, so it gives no additional constraint.

We must pause to discuss an important point, concerning the counting of
boundary conditions versus independent differential equations.  Using the
algebraic constraint (\ref{(tt-ii)}) to eliminate $C_1$,
we have two first order o.d.e.'s for $A_1$ and $B_1$.  On the face of it, our
system looks overconstrained: there are three boundary conditions!
But in fact the system is not overconstrained.  Rather, there is
a constraint on the stress-energy tensor on the 4-brane, due to its
conservation.  By computing $\sum_{a\neq\tr} S^{0a}_{\ \ ;a} \equiv 0$,
where $S_{ab}$ is the surface stress energy on the 4-brane, we obtain
\beq
    {d T_3\over dt} = {\dot B_1\over 2 }\left(T_3(1-\alpha) + T_3'\right)
\eeq
Integrating this, we see that unless the constraint
\beq
\label{stress_constraint}
    T_3' = (\alpha-1)T_3
\eeq
is satisfied, then $T_3$ must have had extra hidden
dependence on $B$ (hence the circumference of the circle
$L_\theta$) beyond that which was explicitly assumed.  If  $T_3$
is truly constant, then any physical model of the stress-energy
must satisfy (\ref{stress_constraint}). This is true for the model
which corresponds to delocalizing  (smearing) a 3-brane around the
circular dimension of the 4-brane, since there $\alpha=1$ and
$T_3'=0$.  And it tells us that the Casimir energy model with
$\alpha=5$ must satisfy $T_3'=4 T_3$, as we saw earlier was indeed
the case for massless particles. With any such choice, it is easy
to see that there are not really two boundary conditions at the
4-brane; rather, imposing the first b.c.\ (\ref{bcA}), together
with the $(tr)$ equation (\ref{(tr)}), implies the second b.c.,
(\ref{bcB}). The result of this discussion is that it suffices to
impose just one b.c.\ at $\tr=\tR$, say (\ref{bcA}), which using
(\ref{stress_constraint}) can be written more simply as
\beq
    A_1' = \frac12 A_0' C_1 -{(\alpha-1)\over 8}(B_0'-A_0') B_1.
\eeq

We solved the above system of equations for the radion numerically,
for the case of $\alpha=1$ (the general result for arbitrary values of
$\alpha$ will be given below) and we
find that it has a negative value of $m^2_r$---it is a tachyon.  The graph of
its dependence on the ratio of warp factors between the 4-brane and the
3-brane is shown in figure 3a.  If we normalize $a(0)=1$ at the 3-brane, then
in the regime where the hierarchy is large (right hand side of the graph),
the radion mass squared depends on $a(\tR)$ ($a$ evaluated at the 4-brane) as
\beq
    m^2_r \cong -20\, L^{-2}\, a(\tR)^{-3/2}\qquad(\alpha=1{\rm\ case\
only}),
\eeq
where we recall that $L$ is the AdS curvature radius.  Since $L\sim 1/$TeV
and $a(\tR)^{3/2} \cong 10^{32}$ to solve the hierarchy problem, we obtain
\beq
\label{radmassval}
    \sqrt{-m_r^2} \sim 10^{-3}{\rm eV}.
\eeq
This is well above the present Hubble scale, so we would have noticed the
expansion or contraction of the extra dimension due to the change in the
strength of gravity, if eq.\ (\ref{radmassval}) were true.

\bigskip
\centerline{\epsfysize=2.1in\epsfbox{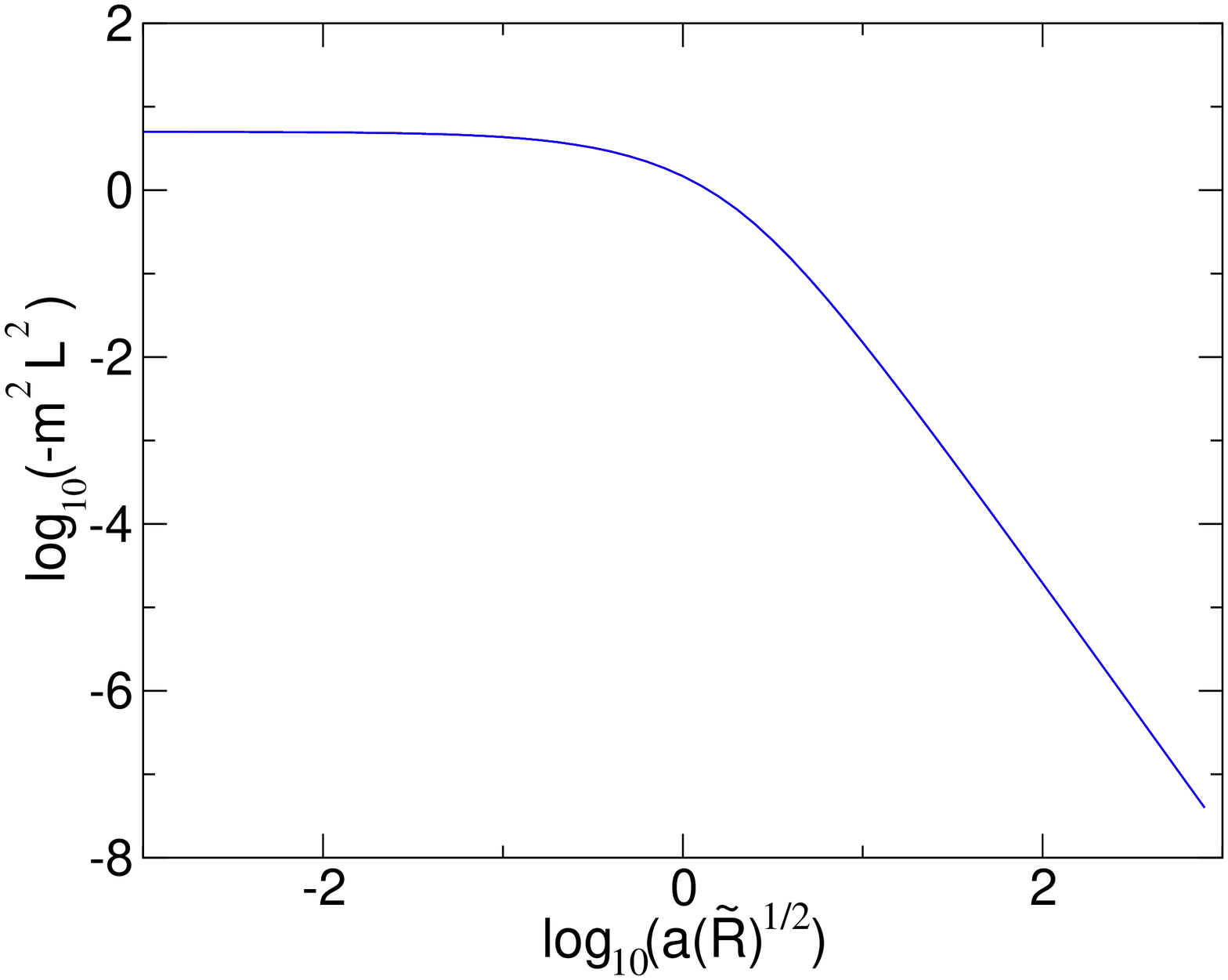}\hfil
\epsfysize=2.45in\epsfbox{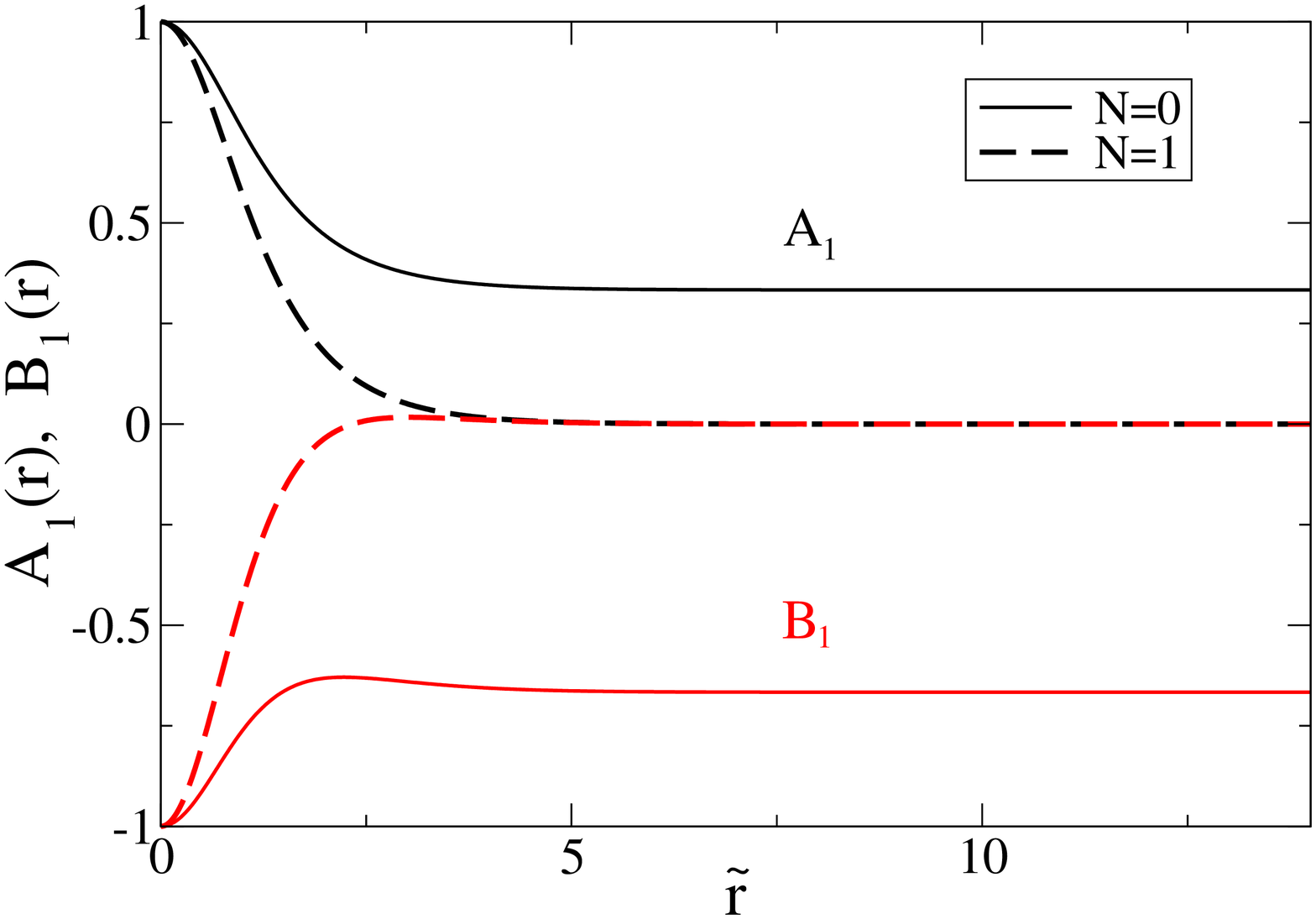}}
\bigskip
\noindent{\small
Figure 3. (a) Log of minus the radion mass squared versus log of the warp
factor, for the case $\alpha=1$.  (b) Outer (solid) lines: the radion wave function for a large
value of the warp factor; inner (dashed) lines: wave function of the first
KK excitation of the radion.}
\vspace{+0.1in}

The radion wave function in the large warp factor regime is shown
in figure 3b.  If the 4-brane is moved even farther away (larger
warp factor), the wave function retains the same form, since it
stays flat in the region of large $\tr$.  Although these plots
were made for the case $\alpha=1$, we find that the wave function
looks essentially the same for all values of $\alpha$.   We see
from its functional form that near the 3-brane $A_1\cong -B_1$,
and since $C_1$ is constrained to be $3A_1+B_1$,  therefore
$C_1\cong -2B_1$ in this region.  For larger values of $\tr$ we
have $A_1=1/3$ and  $B_1=-2/3$, so $C_1$ is extremely small
throughout most of the bulk.  Nevertheless its integral is
nonvanishing, so the radial size of the extra dimension, which is
given by $\int_0^\tR d\tr\, C(\tr)$, changes in response to the
instability, and it does so in the same sense as the size of the
compact dimension, because of our choice of signs in the
defintions of $B_1$ and $C_1$.    That is, the instability is a
simultaneous growth or shrinking of the radius together with the
circumference.   Either direction is a possibility, since the
static solution is analogous to sitting on the top of a hill: the
ball can roll down in either direction.  The situation is
illustrated in figure 4.  Also shown there is the fact that the
relative size of the brane directions, $x^\mu$, grow or shrink in
the opposite sense relative to the extra dimensions.  In the case
where the extra dimensions grow, the endpoint must be the AdS
soliton solution with no 4-brane, since this has been demonstrated
to be the minimum energy solution which  is stable  \cite{spin2}.
In the case where they shrink, the 2-D surface presumably
degenerates into a point.

In comparing the radion in this model to that of the 5-D Randall-Sundrum model,
we can notice several similarities and differences.  Similarly
to the 5-D model  \cite{CGR, CGK},  in 6-D the radion
is an admixture of the radial and brane metric components, such that
oscillations of the radial size are accompanied by
fluctuations in the scale factor of the 4-D universe which are 180$^\circ$
out of phase.  But in 5-D, the radion was exactly massless in the
absence of stabilization, whereas in 6-D it is a tunable parameter.
Another difference is
that, whereas in 5-D the radion has no tower of KK excitations, in 6-D it
does. The mass gap is of order $1/L$, {\it i.e.,} the TeV scale.  The first
few eigenvalues, for large values of the warp factor, are given in table 3.
One notices that these masses are systematically smaller than those of the
graviton and vector modes.  Thus the 

\bigskip
\centerline{$\!\!\!\!\!\!\!\!$
\epsfysize=2.8in\epsfbox{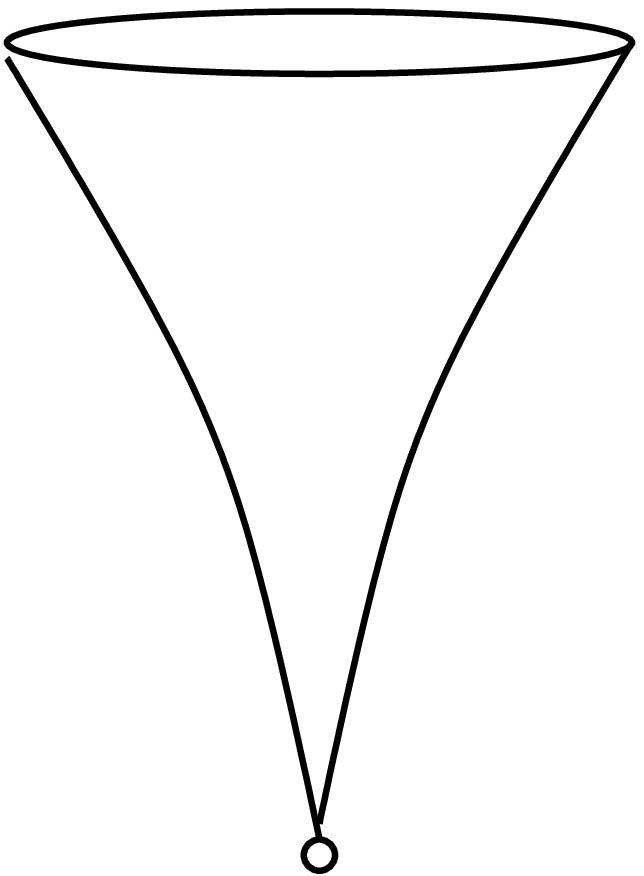}
\epsfxsize=0.5in\epsfbox{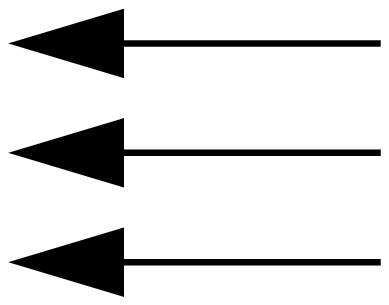}\ \ \ \ 
\epsfysize=1.8in\epsfbox{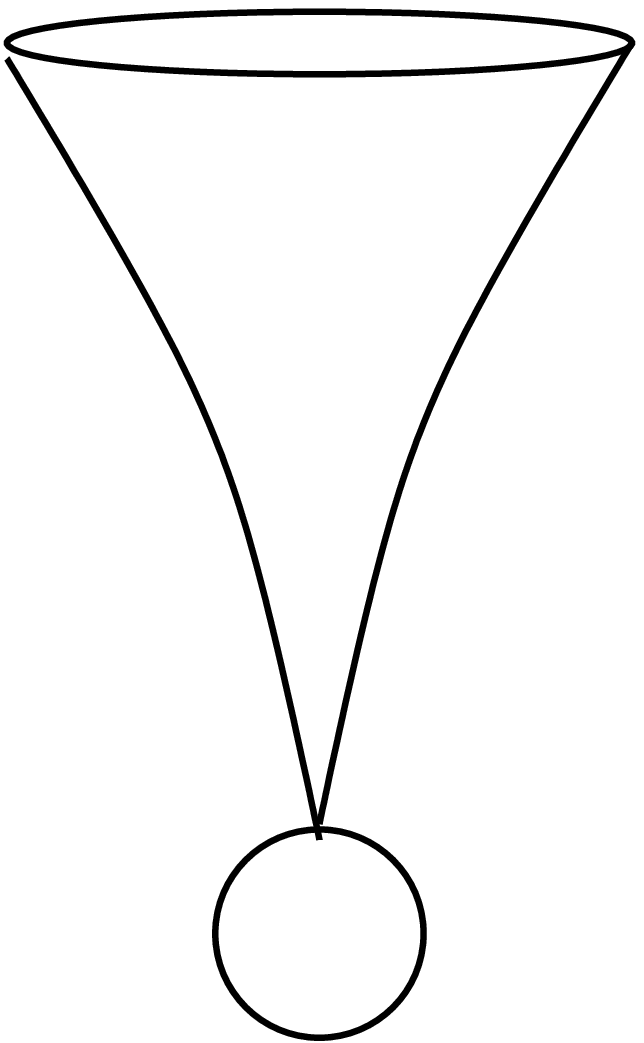}$\!\!\!\!$
\epsfxsize=0.5in\epsfbox{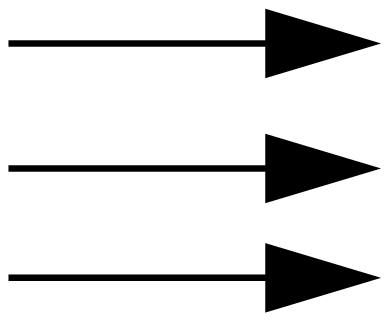}
\ \ \ \ \ \ \ \ \epsfysize=1.1in\epsfbox{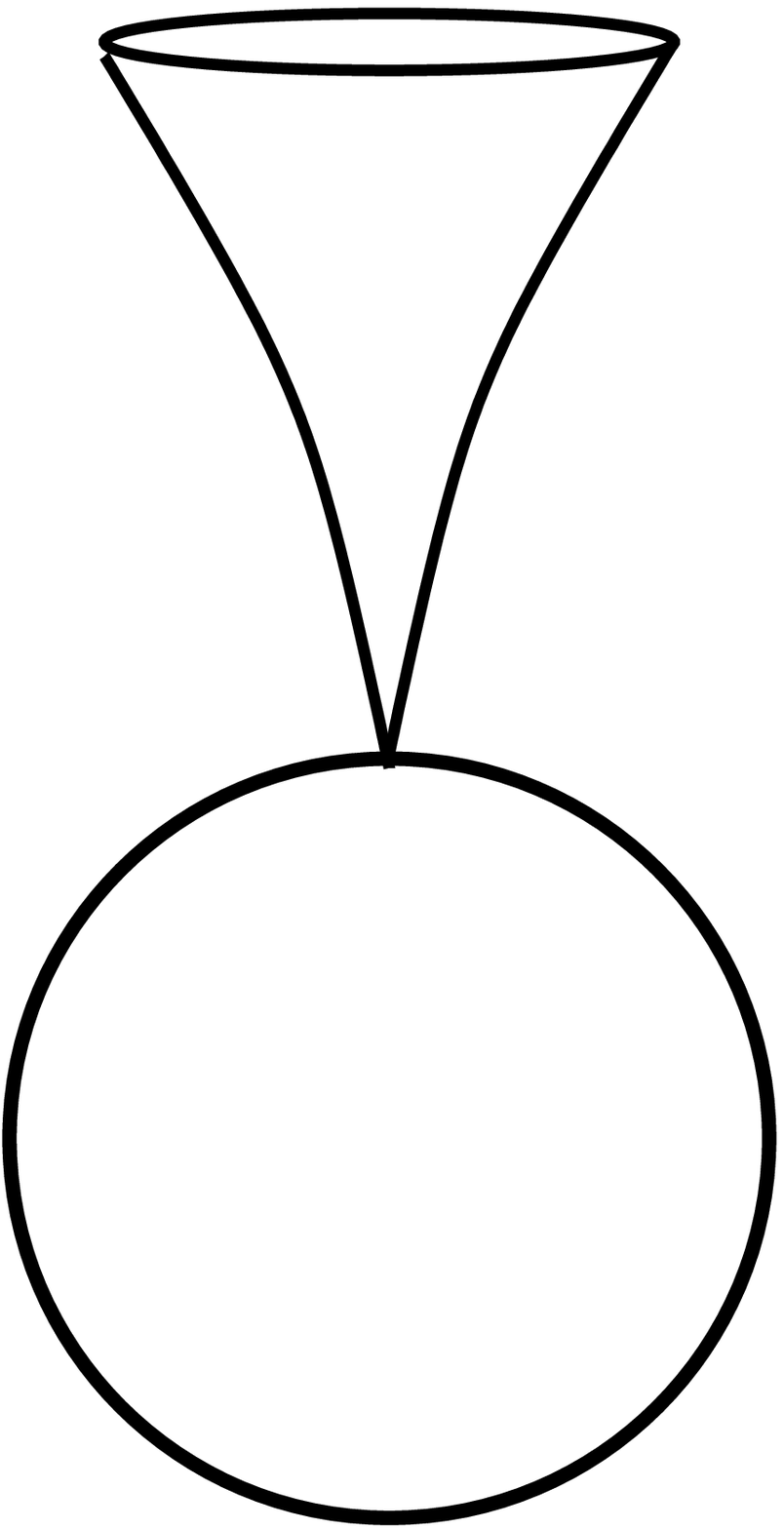}
}
\bigskip
\noindent{\small
Figure 4. Illustration of the instability.  A given static solution, shown in
the center, is unstable toward growth (left) or shrinkage (right) of the
extra dimensions (funnel), accompanied by the opposite behavior of the directions
$x^\mu$ within the 3-brane (shown as a sphere).}
\vspace{+0.1in}

\begin{center}
\begin{tabular}{|c|c|} \hline
Mode Number & $ m_r^2 L^2$ \\   \hline
1 & \p1 1.0188  \\ \hline
2 & \p1 6.1512 \\ \hline
3 & 12.748  \\ \hline
4 & 21.311  \\ \hline
5 & 44.437 \\ \hline
6 & 75.588\\ \hline
\end{tabular}
\end{center}
\noindent{\small
Table 3. Mass squared of the radion KK modes, in units of the
AdS curvature radius, in the limit of large warp factor.}\\

\noindent radion excitations would be the first
signs of new physics from this model (once we have made it viable by
stabilizing the radion) to appear in accelerator experiments. The wave
functions of the second and third excited states are shown in figure 5,
while that of the first excited state appears in figure 3b.  Similarly to
the spin 2 excitations, the modes above the ground state have wave functions
which are exponentially strongly peaked on the TeV brane.  The wave function
of the radion ground state, on the other hand is relatively flat throughout
the bulk.  This plays an important role in its couplings, as we will discuss
in the next section.

We can understand the preceding results for the radion ground state mass and
wave function analytically, and generalize them to arbitrary values of the
4-brane stress-energy parameter $\alpha$.  Toward this end, we first convert the
coupled first order equations into a single second order equation.  The form
of the equations suggests that a natural dependent variable to consider is
the linear combination $H=3A_{1}+B_{1}$.   The variables $A_1$, $B_1$ and
$C_1$ can be expressed in terms of $H$ using eqs.\ (\ref{(tt-ii)}) and
(\ref{(tr)}), where for later convenience we continue to show the effect of
a bulk scalar field, even though we set it to zero for the present:

\centerline{\epsfysize=2.85in\epsfbox{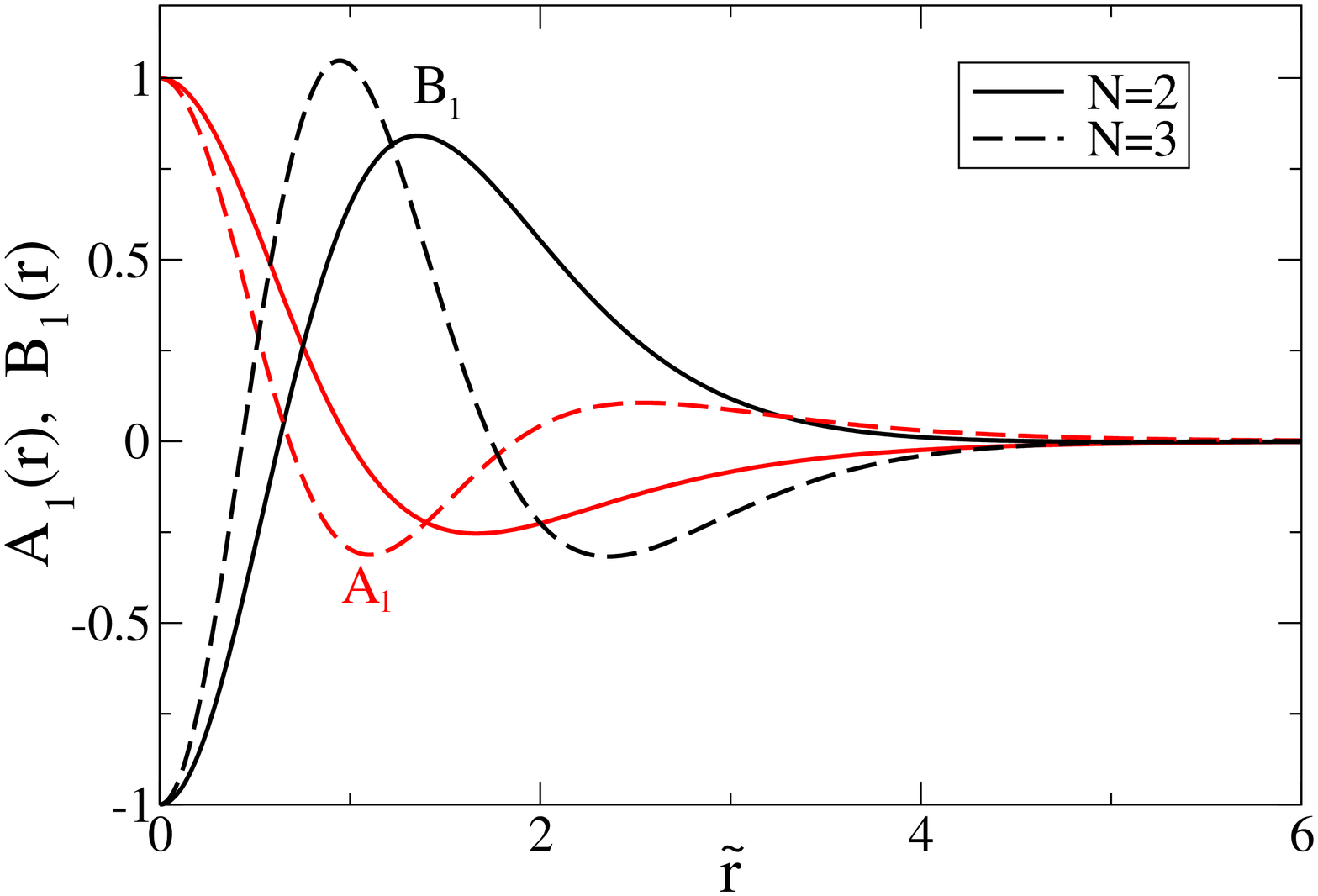}}
\noindent{\small
Figure 5. Solid lines: wave function of second excited state of the radion;
dashed lines: the third excited state.}
\beqa
     A_{1}&=&\frac{1}{2B_{0}'}\left[-H'+\left(A_{0}'+B_{0}'\right)H
     -2\kappa^{2}\phi_{0}'\phi_{1}\right]\nonumber\\
     B_{1}&=&\frac{3}{2B_{0}'}\left[H'-\left(A_{0}' +\frac{1}{3}B_{0}'\right)H
     +2\kappa^{2}\phi_{0}'\phi_{1}\right]\nonumber\\
     C_{1}&=&\frac{1}{2B_{0}'}\left[H'+(B_{0}'-A_{0}')H
     -2\kappa^{2}\phi_{0}'\phi_{1}\right].
\eeqa
Substituting these into the remaining field equation (\ref{(rr)}),
we obtain 
\beqa
    H''+\frac{6A_{0}'^{2}-3B_{0}'^{2}-6A_{0}'B_{0}'-2\kappa^{2}{\phi_{0}'}^{2}}
    {2B_{0}'}H'+\left(\frac{(B_{0}'-A_{0}')(3A_{0}'^{2}+2B_{0}'A_{0}'
    -\kappa^{2}{\phi_{0}'}^{2})}{B_{0}'}
    +\frac{m_{r}^{2}}{a_{0}}\right)H\nonumber\\
    -4\kappa^{2}\left(\frac{\partial V}{\partial \phi}-\kappa^{2}
    \frac{\phi_{0}'}{B_{0}'}V\right)\phi_{1}=0
\eeqa
and the boundary conditions at $\tr=\tR$ and $\tr=0$, respectively
are
\beqa
\left( A_0'+{1-\alpha\over 4}B_0'+
{ 2\kappa^{2}{\phi_{0}'}^{2}\over 3(B_0'-A_0')} \right) { H'} &=& \nonumber\\
 \left( \left(A_0' +
 \frac13 B_0'\right) \left( A_0' +{1-\alpha\over 4}B_0' \right)
\right.\!\!\!\!&-&\!\!\!\!\left.\frac23\kappa^{2}{\phi_{0}'}^{2}
+\frac{2 B_0' m^2_r}{3(B_0'-A_0')a_0}\right) H, \qquad \tr=\tR \nonumber\\
H'&=&A_0' H = 0,\qquad\qquad\qquad\qquad\qquad\!\!\tr =0
\eeqa
The radion mass squared comes into the boundary condition because, in
the process of eliminating $A_1$, $A_1'$, $B_1$ and $B_1'$ in favor of $H'$
and $H'$, it is necessary to use the bulk equation $(\ref{(rr)})$, evaluated
at the 4-brane.

Now it happens that the important features of the radion ground state can
be deduced from approximating the above equations by the form which they
take in the asymptotic region of $\tr\sim\tR$ when $\tR$ is large.
In this region, we have
\beq
    A_0'\cong B_0' \cong -\frac45 k;\qquad B_0'-A_0' \cong
    -8ke^{-2k\tr} \equiv \delta A'
\eeq
and the equations simplify considerably:
\beqa
\label{bulkeq}
&&  H'' - \frac32 A'H' + \left(5A'\delta A' + {m^2_r\over a_0}\right)H
    = 0 \quad{\rm\ in\ the\ bulk};\\
\label{asymbc}
&&\left( {\kappa^2\phi_0'^2\over A'} +\frac38(5-\alpha)
{ \delta A'} \right) { H'} =
\left({\frac {{ m^2_r}}{{ a_0}}}+ \left(\frac{5-\alpha}{2}{ A'}-{
{\kappa^2\phi_0'^2\over A'}}\right) { \delta A'} \right) H
    \quad{\rm\ at\ } \tr=\tR,
 \eeqa
where now we take $A'$ to have the constant value $-4k/5$.
The terms $\delta A'$ and $m^2_r/a_0$ are both of order $e^{-2k\tr}$
in the large $\tr$
region (as we will verify self-consistently),
so that in the bulk equation (\ref{bulkeq}) they can be ignored
compared to the other terms.  The solution in the bulk
has the form
\beq
\label{approxH}
    H(\tr) \cong c_1 + c_2 e^{-(6k/5)\tr} + \delta H,
\eeq
where $\delta H$ represents the small effect of the parenthetical terms
we have ignored.  The latter give rise to a negligible effect on the
bulk solution, $\delta H\ll H$.
However, the small terms proportional to $\delta A'$ and $m^2_r/a_0$
{\it cannot} be neglected
when applying the b.c.\ at the 4-brane.  In fact, this equation,
(\ref{asymbc}), can be used to solve for the radion mass, which in
the absence of the scalar field gives
\beqa
    m^2_r &=& \left. {5-\alpha\over 4} a_0\,\delta A'
\left(\frac32 {H'\over H} - 2A'\right)
    \right|_{\tr=\tR} \cong (\alpha-5){20\over 2^{4/5}} L^{-2} e^{-(6k/5)\tR}
    \nonumber\\
    &=& (\alpha-5){5\over L^2}a_0^{-3/2}(R) + O(a_0(\tR)\delta A'^2)\nonumber\\
     &=& {\alpha-5\over 2}{\Lambda \over a_0^{3/2}(R)} +
    O\left(\Lambda \over a_0^4(R)\right)
\eeqa
The final expression assumes that $a_0$ is normalized to unity at the TeV
brane, and uses the relation (\ref{Lambdaeq}) between $\Lambda$ and $L$.
(The intermediate factor of $2^{4/5}$ comes from
$a_0(\tR)=\cosh^{4/5}(k\tR)$.) Interestingly, the value for $m^2_r$ which we
so obtain is completely insensitive to the details of ${H'\over H }\sim
e^{-(6k/5)\tR}$, much less $\delta H$, since all of these are much smaller
than $A'$.  We  are therefore able to give a very accurate analytic estimate
for the radion mass squared, when the warp factor is large.  The  small
magnitude of $m^2_r$ is seen to be a direct consequence of the value of
$B_0'-A_0'$ in the static solution.  This expression agrees with our previous
numerical results for $\alpha = 1$ (and we have also checked it numerically
for other values of $\alpha$). In the limit that the 4-brane goes to
infinity, so that the full AdS soliton is recovered, the radion becomes
massless, but is not normalizable.  Thus it does not contradict the fact that
the uncut AdS soliton is a stable solution.

Interestingly, the radion mass vanishes almost exactly in the case where the
anistropy of the 4-brane stress tensor is provided by the Casimir  energy and
pressure of fields living on the compact extra dimension. In this case the
relevant energy density scales like $L_\theta^{-5}$, {\it i.e.,} $\alpha=5$, as
expected from dimensional analysis.  This is the unique case where no dimensionful
parameter is introduced in the anistropic part of the 4-brane stress tensor, which
is the part that also controls the position of the 4-brane, and hence how large
the extra dimensions are.  Curiously, the mass does not vanish {\it exactly} when
$\alpha=5$ because of the $O(\delta A'^2)$ correction, whose coefficient turns
out to be $(11+\alpha)/8$.  However, the natural size of this contribution to
$m_r$ is of order $10^{-10}$ eV, which is far below experimental limits on
scalar-tensor theories of gravity.

\section{Radion stabilization and phenomenology}

In this section we show how to increase the mass of the radion through
using a bulk scalar field, and discuss the implications of the model for
collider experiments, tests of the gravitational force, and cosmology.

\subsection{Stabilization by a bulk scalar field}

We have found that the radion can be massive, massless, or tachyonic,
depending on the value $\alpha$ which controls the dependence of the 4-brane
stress-energy on its circumference.  In the latter two cases ($\alpha\le 5$),
it is certainly necessary to increase the radion  mass squared so that we
have a stable universe, with Einstein gravity rather than scalar-tensor
gravity at low energies.   In the 5-D RS1 model, this was achieved by
Goldberger and Wise \cite{GW} by adding a bulk scalar field, whose VEV's at the two
branes were constrained by potentials on the branes to take certain values.
The bulk scalar then acts like a spring between the branes, whose
gradient energy becomes repulsive if the branes get too close, and whose
potential energy (from $m^2\phi^2$) causes attraction if the branes separate
too much.  We expect that the same mechanism should work in 6-D.

Scalar fields in AdS have solutions which are exponentially growing
or decaying toward the ultraviolet cutoff brane. Ref.\  \cite{nelson2} studied
these solutions and found the approximate behavior
\beq
    \phi(\tr) = \phi_+ e^{\sigma_+\tr} +  \phi_- e^{\sigma_-\tr}
\eeq
where $\sigma_\pm = -k\pm \sqrt{k^2+m^2}$.  Near the 3-brane, where the space
does not look like AdS, the behavior is different; $\phi(\tr)\cong
\phi_0(1+m^2\tr^2/4)$, but this will not be very important for understanding
the effect of the scalar since most of the volume of the extra dimensions is
near the 4-brane. For generic boundary conditions, the growing solutions
dominate, and it is a good approximation to neglect the decaying ones.  The
main point is that the most natural configurations are ones where $\phi(0)
< \phi(\tR)$.

Before doing any analytic estimates, we solved the entire system of Einstein
equations numerically, to find the effect of the scalar field on the radion
mass.  There are three kinds of corrections to consider.  First, the scalar
field induces a small back-reaction on the static solutions, $A_0$, $B_0$,
determined by the zeroth order truncation of the Einstein equations
(\ref{tt}-\ref{tr}).  This effect has been analytically computed in
 \cite{nelson2}.  Second, the background scalar configuration couples to the
fluctuations of the metric. This arises solely through the term
$\kappa^2{\phi_{0}'}^{2}C_{1}$ of the  perturbed ($rr$) component of the
Einstein equations, (\ref{(rr)}).  We will see that this is the really
important effect for stabilizing the radion.  The third kind of correction
is from fluctuations of the scalar field, which can mix with the radion.
These are governed by the perturbed scalar field equation (\ref{scalar}).
\beq
   \phi_{1}''-\frac{1}{2}(4A_{0}'+B_{0}')\phi_{1}'-\phi_{0}'H'
    -\frac{1}{2B_{0}'}\frac{\partial V}{\partial \phi}\left((B_{0}'-A_{0}')H+H'
    -2\kappa^{2}\phi_{0}'\phi_{1}\right)
    -\frac{\partial^{2}V}
    {\partial \phi^{2}}\phi_{1}
    +\frac{m_{r}^{2}}{a_{0}}\phi_{1}=0
\eeq
where $V=\frac12 m^2\phi^2$ is the bulk potential.

Our numerical results demonstrating the stabilization of the
radion are shown in figure 6.  We considered scalar field
configurations with $\phi=0$ at the 3-brane and varied the value
of $\phi$ at the 4-brane, showing that $m^2_r$ (in the tachyonic
case $\alpha=1$) becomes positive for sufficiently large values of
$\phi(\tR)$. (Treating $\phi(\tR)$ as an adjustable parameter can
be justified by imagining that we have stiff potentials for $\phi$
on the branes, fixing their boundary values to whatever we
desire.) We checked that these results are quite insensitive to
whether the fluctuation of the scalar are included.  The mixing
between the radion and $\phi_1$ was found to be negligible.

\centerline{\epsfysize=2.85in\epsfbox{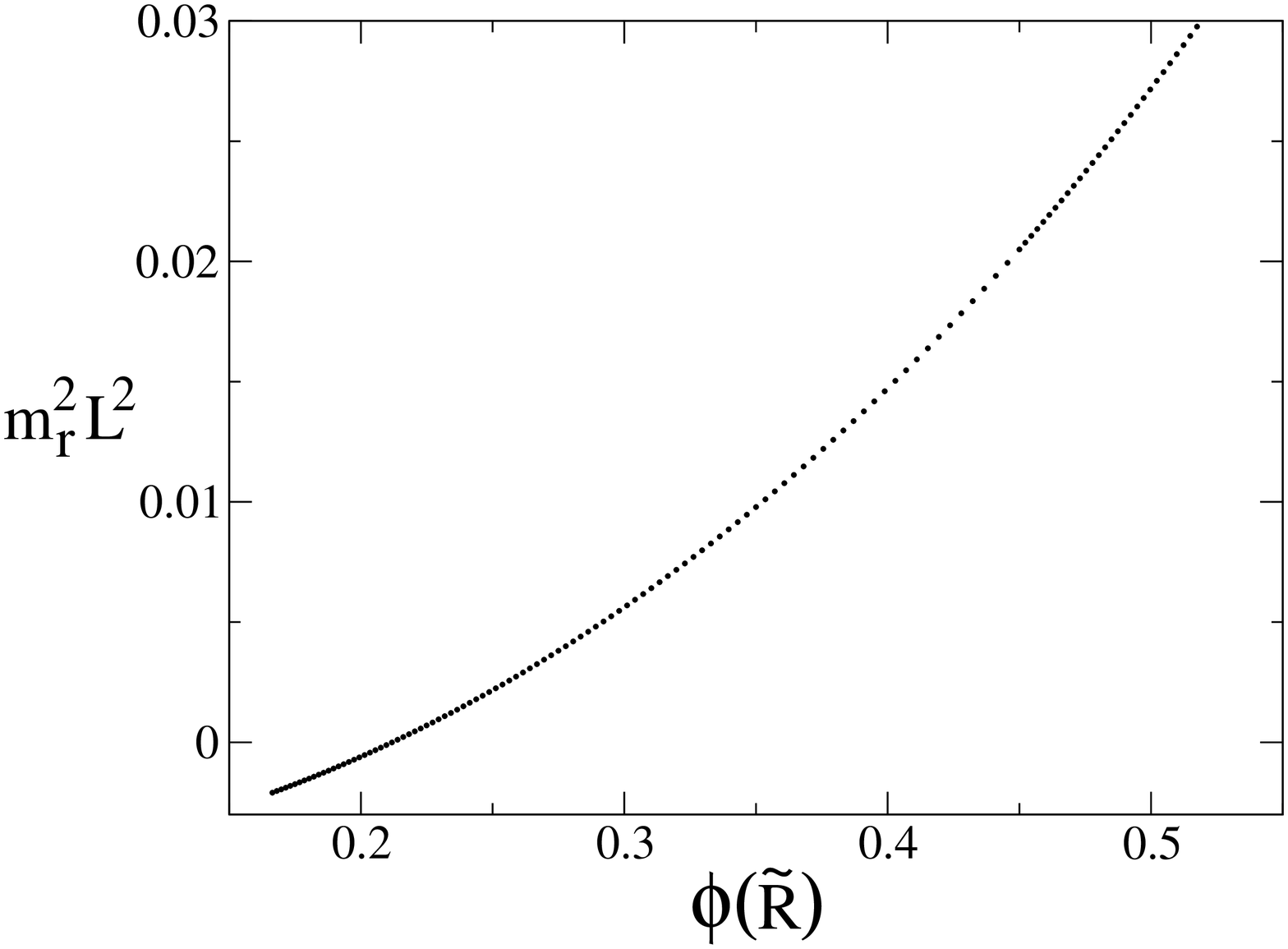}}
\centerline{\small
Figure 6. Dependence of $m^2_r$ on the value of $\phi$ at the 4-brane.}
\vspace{+0.1in}

The behavior shown here can be easily understood by generalizing our previous
derivation of the radion mass to include the effect of the scalar.  The
equation of motion and boundary condition for $H$, in the large-$\tr$ region,
were given in (\ref{bulkeq}-\ref{asymbc}).  We can take
$m^2$ of the scalar to be small, so that its effect can be neglected in the
bulk equation and the approximate solution (\ref{approxH}) is still valid.
Now when we solve the b.c.\ for the radion mass, we obtain the previous
expression plus a new term,
\beq
\label{stab_mass}
    {m^2_r\over a_0} = {5-\alpha\over 4}
    \delta A' \left(\frac32 {H'\over H} - 2A'\right)
    + {H'\over H}{\kappa^2\phi_0'^2\over A'}
\eeq
whose origin can be traced to the extra term in the ($rr$) Einstein equation.
The first term is also changed by the presence of the scalar field, because
of its back-reaction on the static metric.  However, using the results of
ref.\  \cite{nelson2} who computed this back-reaction, we find that $\delta A'$
is still of order $e^{-2k\tR}$.  Therefore, since ${H'\over H}$ is of order
$e^{3k\tr/2}$, the new term on the r.h.s.\ of (\ref{stab_mass})
is the dominant one.  The
fact that $H'\over H$ has the correct sign (negative) to insure that the radion
mass squared is positive is not obvious, but by numerically solving for $H(r)$
we have verified that indeed ${H'\over H}(\tR)<0$, as we show in figure 7.
We thus find that for large enough values of $\phi(\tR)$, the radion mass is
\beq
    m_r \sim {m^2\phi(\tR)\over k M_6^2} e^{-k\tR/5} \sim {\rm\ MeV}
\eeq
independently of the details of the 4-brane stress-energy.

It is remarkable that the stabilized mass is not of order the TeV
scale, as was the case in 5-D RS1 \cite{GW, CGK, MT}. The mass squared is suppressed
by the fractional power of the warp factor left in the product
$a_0 {H'\over H} \sim a_0^{-1/2}$.   Recalling that the Planck
scale hierarchy was set by $a_0^{3/2} \cong 10^{32}$, we see that
the stabilized $m_r$ is suppressed by the factor $10^{16/3}$,
giving $m_r \sim 10$ MeV. This is precisely the same factor by
which a TeV mass particle, transported from $\tr=0$ to $\tr=\tR$,
falls short of the Planck scale, as we noted in section 2.2.  Thus
the smallness of the radion mass seems to be associated with the
additional dilution of the strength of gravity that comes from the
large extra dimension.

\centerline{\epsfysize=2.85in\epsfbox{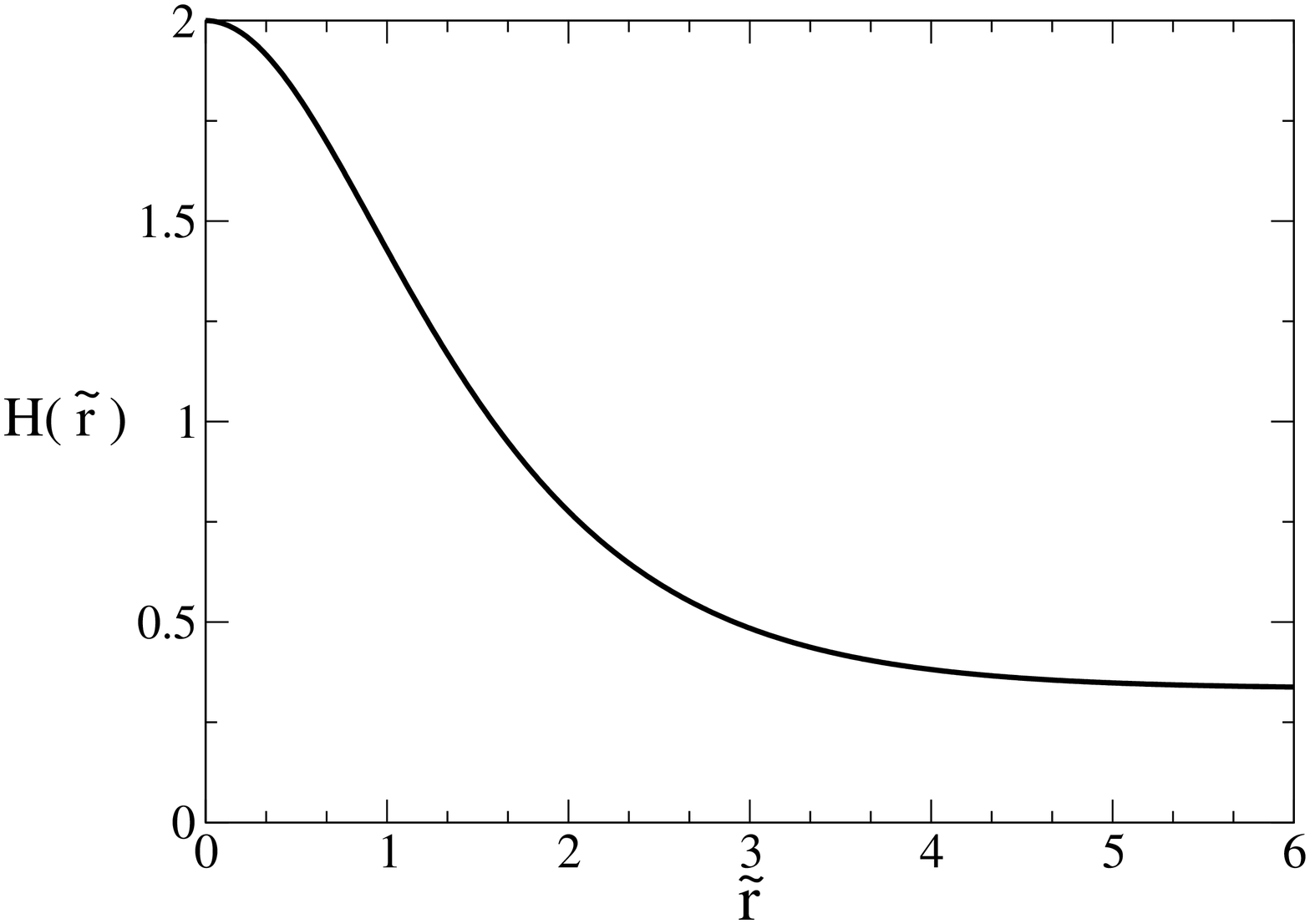}}
\noindent{\small
Figure 7. The radion wave function $H(r)$ for the ground state, showing that
$H'/H < 0$.}
\vspace{+0.1in}

In ref.\  \cite{nelson2}, one advantage of having a bulk scalar field was
already pointed out: because of the back-reaction of the scalar on the
metric, the jump conditions at the 4-brane can generically be satisfied for
large values of $\tR$, as desired for achieving a large hierarchy, without
much sensitivity to the model of stress energy on the 4-brane.  In
particular, choices like $\alpha=0$ (pure tension) or $\alpha=1$ (smeared
3-brane), which by themselves could not have yielded a satisfactory value of
$\tR$, become viable in the presence of the bulk scalar.  The only
requirement for getting sufficiently large $\tR$ is that the scalar mass
should be somewhat light since \beq \tR \sim {k\over m^2}. \eeq Hence one
needs $k/m\sim 8$ to get the desired hierarchy. A similar relation occurs in
the 5-D realization of Goldberger and Wise \cite{GW}. This requirement is quite
compatible with the parameters we need for generating the radion mass, as in
retrospect one would have expected.

\subsection{Couplings of the radion and its excitations}

In the 5-D RS1 model, the stabilized radion has a TeV scale mass and TeV
suppressed couplings to standard model matter \cite{radphen}.  Were the couplings of
our MeV-scale radion so large, it would easily be observable in low-energy
experiments and possibly affect the cooling of supernovae.  Here we show
that the couplings are actually Planck scale suppressed.

Computing the 4-D effective Lagrangian for the gravitational fluctuations
at quadratic order, we obtain
\beq
\label{Leff}
{\cal L}_0 = \frac{1}{3\kappa^2}
\int d\theta\,dr\,a_0(r)\sqrt{b_0(r)}\,
\left[\dot H^2(r,t)+ 2\dot B_1^2(r,t)\right] + A_1(0,t) {T}^\mu_{\ \mu}
 \eeq
where ${T}^\mu_{\ \mu}$ is the trace of the 3-brane stress-energy tensor,
representing the standard model, $H(r,t) = H(r)\varphi_0(t)$, $B_1(r,t) =
B_1(r)\varphi_0(t)$, {\it etc.}\   Here $\varphi_0(t)$ represents the
4-D ground state radion field, and $H(r)$, $B_1(r)$, $A_1(r)$ are the
corresponding wave functions
found in the previous section.  Since they are nearly constant throughout
the bulk, we can take them out of the integral and perform it to obtain
\beq
{\cal L}_0 \sim  M_p^2 \dot\varphi_0^2 +  \varphi_0 {T}^\mu_{\ \mu}
\eeq
This shows that the canonically normalized radion field ground state
has Planck-suppressed couplings to TeV-brane matter.  This differs from
the behavior of the radion in the 5-D RS1 model.  There, the wave function of
the radion is exponentially peaked at the Planck brane, which overcomes
the exponential warp factor in the measure to give ${\cal L}_0\sim $ (TeV)$^2
\dot\varphi_0^2 +  \varphi_0 {T}^\mu_{\ \mu}$ instead.  The flatness of the radion
wave function in the present case accounts for its weak couplings to the
TeV brane.

The KK excitations of the radion {\it are} exponentially peaked on
the TeV brane, on the other hand.  We can understand this from the asymptotic
form of the bulk equation of motion (\ref{bulkeq}); since the mass is no
longer negligible, the solutions behave like $H(r) \cong c_2 e^{-6k\tr/5}$,
with the constant piece $c_1$ equal to zero.
The integrand of (\ref{Leff}) behaves like $e^{-12k\tr/5 + 12k\tr/5}
= O(1)$, so we obtain
\beq
{\cal L}_n \sim  {M_6^4\over k} \tR\, \dot\varphi_n^2 +
\varphi_n {T}^\mu_{\ \mu}\qquad\to\qquad \dot\varphi_n^2 + M_6^{-2}\sqrt{k\over\tR}\
    \varphi_n {T}^\mu_{\ \mu}
\eeq
Hence the coupling of the radion excited state is suppressed only by the
small factor $(M_6\tR)^{1/2}\sim \sqrt{60}$ relative to the TeV scale.
The radial KK gravitons have similar couplings, but larger masses
(compare Tables 1 and 3), so the radion excitations would be the first signal
of new physics in collider experiments.  Heavy radions could be copiously
produced in the $s$-channel at the LHC, through gluon-gluon fusion events
due to the QCD trace anomaly contribution to ${T}^\mu_{\ \mu}$.

\subsection{Gravity and cosmology}

Although perhaps less physically motivated, models of the 4-brane
stress-energy with $\alpha>5$ predict that the radion mass squared will be
positive even without a bulk scalar, and that its magnitude is in the
$10^{-3}$ eV regime.  This is within the reach of Cavendish-type tests of
submillimeter gravity  \cite{adelberger}, and  will be even more accessible to
upcoming  versions of the experiment which will have improved sensitivity.

It may also be possible to achieve this situation without appealing to
exotic forms of matter on the 4-brane.  The radion mass will get radiative
corrections from its couplings to matter.  Since the radion couples to
the trace of the stress energy tensor on either brane, the heaviest 
particles will contribute the most strongly.  Considering matter which is on the
TeV brane, we can estimate the size of the one-loop correction as
\beq
	m^2_{r,{\rm 1-loop}} \sim { {\rm TeV}^4\over M_p^2}
\eeq
The numerator comes from the fact that the TeV scale is the cutoff on
the 3-brane, and the heaviest particles will have masses of this order,
whereas the denominator is due to the fact that the lowest mode of the
radion has Planck-suppressed couplings.  This argument could be upset
if the 4-brane has massive particles which are much heavier than the
TeV scale, since by the same argument these could apparently make the
radion very heavy and presumably would destabilize the hierarchy which
we have achieved.  It may be necessary to assume that there are only 
massless particles on the 4-brane to avoid this.

On the other hand, if we allow heavy particles to exist on the 4-brane,
their natural mass scale is $\sqrt{a(R)}$ TeV $\sim 10^{13}$ GeV.  It is
interesting that this is the right order of magnitude for generating the
obsserved primordial density fluctuations from the simplest model of chaotic
inflation.  This is an advantage of the present model over the 5-D RS model,
where  a $\sim 10^{13}$ GeV would look unnaturally light were it living on
the Planck brane, and of course too heavy to exist on the TeV brane.

\section{Summary}

We have focused on the simplest and most direct generalization to six
dimensions of the 5-D Randall-Sundrum two-brane model: the AdS soliton model,
with the TeV 3-brane at the center of the azimuthally symmetric extra
dimensions, and a 4-brane cutting the space off at some finite radius. The
model has many features in common with its 5-D predecessor: the geometry is
highly warped and very close to AdS in the region far from the 3-brane, the
graviton zero-mode is localized on the hidden brane, while the radial KK
excitations are localized on the TeV brane and have a TeV mass gap.  In both
models, the radion can easily be stabilized by the Goldberger-Wise mechanism,
using a bulk scalar field.

However, there are also some quite distinctive differences.  The hierarchy
between the Planck and weak scales, while generated mostly ($2/3$) by
warping, is also partly ($1/3$) due to the exponentially large size of the
compact extra dimension  \cite{explarge},
 giving it some features in common with the large
extra dimension proposal.  The mass scale at the 4-brane is not the Planck
scale, but it is suppressed by the size of the large compact dimension to
the $10^{13}$ GeV scale.  There is a tower of relatively light  ($\sim$ TeV)
KK gravitons corresponding to this large dimension.   In the absence of
stabilization by a scalar field, the 6-D model requires some mildly exotic
form of stress energy on the hidden brane in order to have a finite volume. 
The 4-brane stress tensor generically depends on the size of the extra
dimension as $L_\theta^{-\alpha}$ with some model-dependent number $\alpha$.

Most of these features were already known; in the present work we computed
the spectrum of metric perturbations, including the graviton and
graviphoton modes, and we found the unexpected new result that the radion is
not necessarily massless, but has a mass squared which depends linearly on
$\alpha$ and the negative bulk cosmological constant: $m^2_r \sim
(5-\alpha)\Lambda$ TeV$/M_p$.  Only for the special case of Casimir energy
on the 4-brane ($\alpha=5$) is it massless.  For smaller values of $\alpha$
it is tachyonic, and the spacetime is unstable.  Its couplings to the TeV
brane are Planck suppressed rather than TeV suppressed, due to the different
behavior of its wave function relative to the 5-D case.  Once stabilized by
a bulk scalar  field, the radion mass is not TeV scale, as in 5-D, but
rather at the  MeV scale.  This suppression is related to the presence of
the large extra dimension which does not feature in 5-D.

The 6-D model has similar phenomenology to the 5-D model, since the
Kaluza-Klein excitations of the radion behave much like the ground state of
the stabilized 5-D radion.  However, there is a new possibility that the
radion is stabilized not by the Goldberger-Wise mechanism, but by some form of stress
energy on the 4-brane which has $\alpha>5$, or perhaps by radiative corrections
from standard model particles on the TeV brane.  In this case the radion mass
is in the milli-eV range, which is just right for being accessible
in experiments which test gravity below 1 millimeter.

The latter possibility would seem to require the absence of massive particles
on the 4-brane, since radiative effects there should induce much larger 
corrections to the radion mass.  In fact it might be necessary to forbid
heavy particles on the 4-brane just to maintain the large hierarchy we
set out to achieve.  This is a question which deserves further study.  
But if it is consistent to have heavy fields on the 4-brane, then the fact
that their mass is naturally of order $10^{13}$ GeV is intriguing for 
inflation, since this is the right scale for getting density perturbations
of order $10^{-5}$ in chaotic inflation.

We have left for future work a study of 3-brane fluctuation modes, where
the position of the 3-brane could oscillate with respect to the center of
the extra dimensions.  These modes, if they exist and are sufficiently
light or strongly coupled, could be important for the phenomenology of this
model, since they might induce a coupling between the graviphoton and
standard model particles.

{\bf Acknowledgments}: This  research was supported by NSERC of Canada and
Fonds FCAR du Quebec. The research of N.R.C.\ was supported in part by
the D.O.E.\ under cooperative research agreement $\#$ DF-FC02-94ER40818.
We thank Robert Myers, Shinji Mukohyama
and Fr\'ederic Leblond for fruitful discussions.  J.C.\ thanks Nima
Arkani-Hamed for helpful comments and the Harvard theory group for its
hospitality during the completion of this work. N.R.C.\ would like to thank
\O yvind Tafjord for valuable discussions at an early stage of this research.

\end{document}